%% file: main.tex
\documentclass[twocolumn]{aastex631}
\usepackage{CJK}
\usepackage[T1]{fontenc}
\usepackage{booktabs}
\usepackage{longtable}
\usepackage{amsmath}
\usepackage{array}
\usepackage{enumitem}
\usepackage{appendix}
\usepackage{xcolor}
\usepackage{cleveref}
\crefname{section}{sect.}{sects.}
\Crefname{section}{Sect.}{Sects.}
\crefname{figure}{Figure}{figs.}
\Crefname{figure}{Fig.}{Figs.}
\crefname{table}{table}{tables}
\Crefname{table}{Table}{Tables}

\makeatletter
\usepackage{etoolbox}
\patchcmd\H@refstepcounter{\protected@edef}{\protected@xdef}{}{}
\makeatother

\newcommand{\program}[1]{\textsc{#1}}

\begin{document}

\title{SN\,2021aaev: a Hydrogen-Rich Superluminous Supernova with Early Flash and Long-Lived Circumstellar Interaction in an Unusual Host Environment}
\begin{CJK*}{UTF8}{gkai}

\author[0000-0002-9744-3910]{Yang Hu (胡阳)}
\affiliation{The Oskar Klein Centre, Department of Astronomy, Stockholm University, AlbaNova, SE-10691 Stockholm, Sweden}

\author[0000-0001-9454-4639]{Ragnhild Lunnan}
\affiliation{The Oskar Klein Centre, Department of Astronomy, Stockholm University, AlbaNova, SE-10691 Stockholm, Sweden}

\author[0000-0002-8041-8559]{Priscila J. Pessi}
\affiliation{The Oskar Klein Centre, Department of Astronomy, Stockholm University, AlbaNova, SE-10691 Stockholm, Sweden}

\author[0000-0001-8419-3062]{Alberto Saldana-Lopez}
\affiliation{The Oskar Klein Centre, Department of Astronomy, Stockholm University, AlbaNova, SE-10691 Stockholm, Sweden}

\author[0000-0001-8005-4030]{Anders Jerkstrand}
\affiliation{The Oskar Klein Centre, Department of Astronomy, Stockholm University, AlbaNova, SE-10691 Stockholm, Sweden}

\author[0000-0003-1546-6615]{Jesper Sollerman}
\affiliation{The Oskar Klein Centre, Department of Astronomy, Stockholm University, AlbaNova, SE-10691 Stockholm, Sweden}

\author[0000-0001-6797-1889]{Steve Schulze}
\affiliation{Center for Interdisciplinary Exploration and Research in Astrophysics (CIERA), Northwestern University, 1800 Sherman Ave, Evanston, IL 60201, USA}

\author[0000-0003-0227-3451]{Joseph P. Anderson}
\affiliation{European Southern Observatory, Alonso de C\'ordova 3107, Casilla 19, Santiago, Chile}

\author[0000-0003-1325-6235]{Se\'an J. Brennan}
\affiliation{Max-Planck-Institut f\"ur extraterrestrische Physik, Giessenbachstrasse 1, 85748 Garching, Germany}
\affiliation{The Oskar Klein Centre, Department of Astronomy, Stockholm University, AlbaNova, SE-10691 Stockholm, Sweden}

\author[0000-0001-8082-5296]{Stefano P. Cosentino}
\affiliation{University of Catania, Department of Physics and Astronomy ``E. Majorana'', Italy}

\author[0000-0002-3884-5637]{Anjasha Gangopadhyay}
\affiliation{The Oskar Klein Centre, Department of Astronomy, Stockholm University, AlbaNova, SE-10691 Stockholm, Sweden}

\author[0009-0000-9383-2305]{Anamaria Gkini}
\affiliation{The Oskar Klein Centre, Department of Astronomy, Stockholm University, AlbaNova, SE-10691 Stockholm, Sweden}

\author[0000-0002-1650-1518]{Mariusz Gromadzki}
\affiliation{Astronomical Observatory, University of Warsaw, Al. Ujazdowskie 4, 00-478 Warszawa, Poland}

\author[0000-0001-8587-218X]{Matthew J. Hayes}
\affiliation{The Oskar Klein Centre, Department of Astronomy, Stockholm University, AlbaNova, SE-10691 Stockholm, Sweden}

\author[0000-0002-3968-4409]{Cosimo Inserra}
\affiliation{Cardiff Hub for Astrophysics Research and Technology, School of Physics \& Astronomy, Cardiff University, Queens Buildings, The Parade, Cardiff, CF24 3AA, UK}

\author[0000-0003-3939-7167]{Tom\'as E. M\"uller-Bravo}
\affiliation{School of Physics, Trinity College Dublin, The University of Dublin, Dublin 2, Ireland}
\affiliation{Instituto de Ciencias Exactas y Naturales (ICEN), Universidad Arturo Prat, Chile}

\author[0000-0002-2555-3192]{Matt Nicholl}
\affiliation{Astrophysics Research Centre, School of Mathematics and Physics, Queens University Belfast, Belfast BT7 1NN, UK}

\author[0000-0003-0006-0188]{Giuliano Pignata}
\affiliation{Instituto de Alta Investigaci\'on, Universidad de Tarapac\'a, Casilla 7D, Arica, Chile}

\author[0000-0003-2091-622X]{Avinash Singh}
\affiliation{The Oskar Klein Centre, Department of Astronomy, Stockholm University, AlbaNova, SE-10691 Stockholm, Sweden}

\author[0000-0003-0733-2916]{Jacob L. Wise}
\affiliation{Astrophysics Research Institute, Liverpool John Moores University, 146 Brownlow Hill, Liverpool L3 5RF, UK}

\author[0000-0003-1710-9339]{Lin Yan}
\affiliation{Caltech Optical Observatories, California Institute of Technology, Pasadena, CA 91125, USA}

\author[0009-0006-7265-2747]{Judy Adler}
\affiliation{IPAC, California Institute of Technology, 1200 E. California Blvd, Pasadena, CA 91125, USA}

\author[0000-0002-1066-6098]{Ting-Wan Chen}
\affiliation{Graduate Institute of Astronomy, National Central University, 300 Jhongda Road, 32001 Jhongli, Taiwan}

\author[0000-0001-9152-6224]{Tracy X. Chen}
\affiliation{IPAC, California Institute of Technology, 1200 E. California Blvd, Pasadena, CA 91125, USA}

\author[0000-0002-5619-4938]{Mansi M. Kasliwal}
\affiliation{California Institute of Technology, 1200 E. California Boulevard, Pasadena, CA 91125, USA}

\author[0000-0001-6540-0767]{Thallis Pessi}
\affiliation{European Southern Observatory, Alonso de C\'ordova 3107, Casilla 19, Santiago, Chile}

\author[0000-0003-1450-0869]{Irene Salmaso}
\affiliation{INAF–Osservatorio Astronomico di Capodimonte, Salita Moiariello 16, 80131 Napoli, Italy}
\affiliation{INAF-Osservatorio Astronomico di Padova, Vicolo dell'Osservatorio 5, 35122 Padova, Italy}

\author[0000-0002-1229-2499]{David R. Young}
\affiliation{Astrophysics Research Centre, School of Mathematics and Physics, Queen's University Belfast, Belfast BT7 1NN, UK}

\begin{abstract}

We present photometric and spectroscopic observations of SN\,2021aaev, a hydrogen-rich, superluminous supernova with persistent (at least $\sim100$ days) narrow Balmer lines (SLSN-IIn) at redshift $z=0.1557$. We observed SN\,2021aaev to rise in $32.5 \pm 1.0$ days since first light and reach a peak absolute magnitude of $-21.46 \pm 0.01$ in the ATLAS $o$ band. The pre-peak spectra resemble those of typical SNe IIn with flash-ionization features arising from the interaction with a dense, confined circumstellar medium (CSM), albeit the flash timescale is longer than usual ($>20$ days). Post peak, the narrow emission lines evolve slowly, and the absence of ejecta features indicates strong deceleration by the CSM. The total radiated energy (about $1.41\times10^{51}$~ergs) is possible with a low-mass (1--$2\,M_{\odot}$) ejecta ploughing into a massive (9--$19\,M_{\odot}$), extended (outer radius $>1\times10^{16}$~cm) H-rich CSM, or alternatively, with magnetar-powered models. Interestingly, the host environment consists of a spiral galaxy with a red substructure in the south-eastern part, and the SN's exact location coincided with the quiescent red substructure (star-formation rate$=0.02^{+0.13}_{-0.02}\,M_{\odot}$~yr$^{-1}$). Given the atypical environment and the obscuring effect of the massive CSM, a thermonuclear (Type Ia-CSM) origin cannot be ruled out. Altogether, SN\,2021aaev is a compelling case to study the diversity of SLSN-IIn features and their host environment.

\end{abstract}


\section{Introduction}
\end{CJK*}

Superluminous supernovae (SLSNe) are a rare class of extremely luminous stellar explosions, with peak luminosities typically exceeding what can be explained in the framework of ordinary core-collapse supernovae (CCSNe). The exact definition of this class often involves a peak luminosity threshold. Historically, this threshold has been chosen somewhat arbitrarily based on a nominal value (e.g. $-21$~mag in the optical, \citealt{2012Sci...337..927G}; $-20$~mag in the optical, \citealt{Perley2016}) that can separate SLSNe from ordinarily bright SNe. This threshold has evolved over time, with proposed values varying across different subclasses of SLSNe (see e.g. \citealt{2019ARA&A..57..305G, Kangas, pessi2025}).

SLSNe are broadly divided into Type I superluminous supernovae (SLSNe-I; hydrogen-poor SLSNe) and Type II superluminous supernovae (SLSNe-II; hydrogen-rich SLSNe), based on the absence or presence of hydrogen in their spectra. With a growing number of observations, most SLSNe-I have been found to exhibit a series of distinctive \ion{O}{2} features spanning $3500-5000$ \AA\ in their early spectra \citep{2011Natur.474..487Q, 2018ApJ...855....2Q,10.1093/mnras/stw512, 2019ARA&A..57..305G} with a few exceptions \citep[see e.g.][]{2020wnt, Schulze_ibb}. However, for SLSNe-II, no ubiquitous spectroscopic property has been identified. The majority exhibit narrow H lines and can be further classified as SLSNe-IIn (or simply SNe IIn; \citealt{Kangas}), while a smaller subset exhibit broad Balmer emission in their early photospheric phase \citep[see e.g.][]{2008es_2, 10.1093/mnras/stx3179, Kangas}. Some works advocate a more restrictive definition that requires the presence of broad H$\alpha$ emission (FWHM $\geq5000$\,km\,s$^{-1}$) in early photospheric phase for classification as a SLSN-II \citep[e.g.][]{10.1093/mnras/stx3179, Kangas}. However, early-time spectroscopy is not always available, making it difficult to apply such a criterion consistently. In this work, we refer to all H-rich SNe with peak luminosity brighter than $-20$~mag as SLSNe-II \citep[see e.g.][]{pessi2025}, and refer to those with narrow H lines in the early photospheric phase as the subclass SLSNe-IIn.

SLSNe-II are rare. Early studies estimated their occurrence rate to be approximately $0.1\%$ of all CCSNe \citep{Quimby2013, Taylor_2014}. Nevertheless, several individual SLSNe-II have been investigated (e.g. SN\,2006gy\footnote{Some interpret SN\,2006gy as a Type Ia SN interacting with CSM (Ia-CSM), see \cite{2006gy_3}.}, \citealt{2006gy_0, 2007ApJ...666.1116S}; SN\,2008am, \citealt{2008am}; SN\,2008es, \citealt{2008es_1,2008es_2}; SN\,2010jl, \citealt{Stoll_2011, Fransson_2014}; SN\,2016aps, \citealt{Suzuki_2021}; SN\,2017hcc, \citealt{Moran_2023}; SN\,2021adxl, \citealt{Brennan_2024}), along with analyses of small samples (e.g. \citealt{10.1093/mnras/stx3179}). Recently, \cite{Kangas} and \cite{pessi2025} presented studies on the light curves of SLSNe-II from the Zwicky Transient Facility (ZTF) Survey \citep{2019PASP..131a8002B, Masci_2019_ZTF, Graham_2019, Dekany_2020}. \cite{Kangas} studied 14 SLSNe-II with broad Balmer emission in the ZTF sample and \cite{pessi2025} studied the remaining 107 that exceeds the $-20$~mag threshold, and this constitutes the largest SLSN-II light-curve sample to date. These studies show that SLSNe-II have heterogeneous light-curve  properties. They are also energetic events with peak bolometric luminosities typically in the range of $10^{43}-10^{44}$~erg~s$^{-1}$ and can radiate energy greater than $10^{51}$~erg in total. In the extreme cases, the total energy may exceed what can be explained by the core-collapse neutrino-driven explosion mechanism \citep[see e.g.][]{Neutrino1, Neutrino2}. 

Spectroscopically, the majority of SLSNe-II are SLSNe-IIn, and the narrow H lines originate from the photo-ionized, H-rich, slow-moving circumstellar medium (CSM; \citealt{IIn2}). When the SN ejecta collides with this CSM, they generate shock waves that convert ejecta kinetic energy into radiation and power the extraordinarily luminous light curves \citep{Chevalier1994}. Additional spectroscopic evidence for CSM interaction is the presence of highly ionized narrow lines (e.g. \ion{He}{2} $\lambda4686$). These so-called flash-ionization features typically appear within hours to days after explosion, when the ejecta encounters dense and confined CSM, and disappear before peak luminosity \citep[transients exhibiting these features are thus dubbed ``flashers'';][]{Gal_Yam_2014, Khazov_2016, Bruch_2023}. Capturing flash features requires early-time spectroscopy, which remains rare for SLSNe-IIn because surveys (e.g. ZTF) preferentially select SLSN candidates with long rise times to reduce non-SLSN contamination \citep{Kangas,Chen_2023}.

CSM interaction is a promising mechanism to explain both the high peak luminosity and the observed spectral features of SLSNe-IIn. While we directly observe the radiation from CSM interaction, the nature of the progenitor itself is largely obscured by it. In some cases, such as SN\,2006gy, a thermonuclear explosion inside a massive common envelope has been proposed \citep[e.g.][]{2006gy_3}, suggesting that other explosion mechanisms may underlie transients that we perceive as SLSNe-IIn. Meanwhile, the origin of the CSM from the progenitor's mass loss remains an open question, with viable processes such as stellar winds \citep{1970ApJ...159..879L, 2008A&ARv..16..209P}, eruptive mass loss \citep{2003ApJ...591..288H, Woosley2007, 2014ApJ...785...82S, 2017ApJ...836..244W}, or mass transfer in binary or multiple systems \citep{2005A&A...435.1013P, Yoon_2017, binary, 2020A&A...637A...6L}. 

Notably, past studies have already widely used CSM interaction to explain the powering of normal Type IIn supernovae (SNe IIn; \citealt{IIn1}), where the brighter end of the population can reach a peak luminosity $<-19$~mag \citep{Nyholm2020}. The potential overlap with the nominal $-20$~mag threshold has led to questions on the validity of using a magnitude cutoff for classification \citep[see e.g.][]{hiramatsu2024typeiinsupernovaei}, and perhaps, two broader questions regarding the nature of the SLSN-IIn class:
\begin{enumerate}[itemsep=0pt, parsep=0pt, topsep=0pt]
  \item Are SLSNe-IIn simply the luminous extreme of SNe IIn?
  \item What CSM properties, if any, distinguish the higher luminosity of SLSNe-IIn/``luminous'' SNe IIn from those of normal SNe IIn?
\end{enumerate}

In addition to CSM interaction, several other powering mechanisms have been proposed to explain the light curves of SLSNe, but with less observed spectroscopic evidence. While initially suggested for SLSNe-I, these mechanisms can also be applied to SLSNe-IIn and they can work in conjunction with CSM interaction. One explanation involves the spin-down of highly magnetized neutron stars called magnetars as an additional central-engine energy source to power the light curve \citep{1971ApJ...164L..95O, magnetar2, magnetar3}. Another central-engine-powered source is the accretion of fallback material onto a black hole \citep{2013ApJ...772...30D, accretion2}. A third mechanism involves extremely massive progenitors ($M\sim140$--$260\,M_{\odot}$) undergoing thermonuclear explosions triggered by electron-positron pair production known as pair instability supernovae \citep[PISNe;][]{PairInstability,2003ApJ...591..288H, RevModPhys.74.1015, Schulze_ibb}.

Unraveling the mysteries surrounding SLSNe-IIn requires detailed comparisons between theoretical models and observations. To that end, comprehensive case studies of individual SLSNe-IIn are still essential. In this paper, we present a large photometric and spectroscopic dataset for SN\,2021aaev, followed by a detailed analysis. SN\,2021aaev is a SLSN-IIn with peak optical luminosity $<-21$~mag. We observed blended, highly ionized, flash features during the rise of its light curve, and we observed Lorentzian-winged hydrogen Balmer lines throughout all phases. We invoke models with massive and extensive CSM to explain the observed light curve and spectral features. After the work by \cite{Kangas_2025} on the SLSN-II flasher with broad Balmer emission SN\,2023gpw, SN\,2021aaev is the first SLSN-IIn to exhibit flash-ionization features, providing a unique opportunity to explore the diversity within the SLSNe-IIn class. 

This paper is organized as follows. In \Cref{sec:obs}, we present the photometric and spectroscopic data of SN\,2021aaev. In \Cref{sec:phot}, we analyze the multi-band light curves and compare with a SLSN-II sample and other extensively studied luminous H-rich SNe. In \Cref{sec:spec}, we investigate the spectral features, focusing on the Lorentzian-winged narrow H lines and the flash features. In \Cref{sec:model}, we model the light curve with common SLSN models. In \Cref{sec:host_environment}, we analyze the host environment using the host spectral energy distribution (SED) and absorption lines. In \Cref{sec:dis}, we discuss the CSM properties likely responsible for the superluminous nature of SN\,2021aaev and consider SN\,2021aaev's possible relation to other transient types. Finally, \Cref{sec:con} summarizes the key findings of the paper.

\section{Observations} \label{sec:obs}

\subsection{Discovery and classification}

\begin{figure*}[t]
    \centering
    \includegraphics[width=\linewidth]{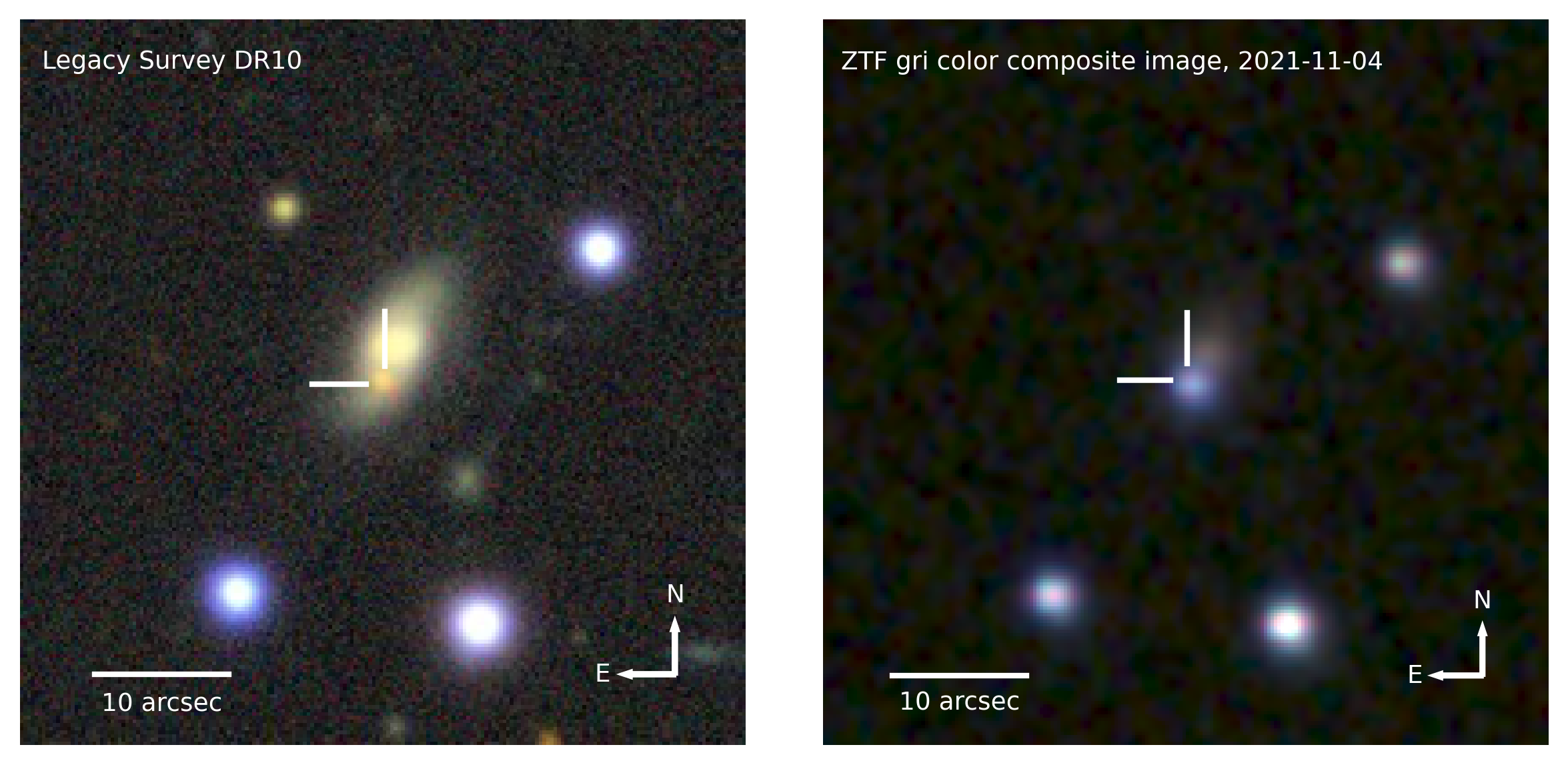}
    \caption{Image of the field of SN\,2021aaev. \textit{Left}: Legacy Survey DR10 \textit{grz} color composite image of the field of SN\,2021aaev before explosion. Interestingly, the location of the SN coincides with a red substructure in the southern part of the host galaxy. We discuss the nature of this red substructure in \Cref{sec:host_environment}. \textit{Right}: ZTF \textit{gri} color composite image of the field of the SN near peak luminosity. The SN has a blue color near peak.}
    \label{fig:field}
\end{figure*}

The discovery of SN\,2021aaev was first reported on the Transient Name Server (TNS) from the ZTF survey as ZTF21aceqrju \citep{DiscoveryTNS} at $20.19$~mag in the $r$ band on Oct 10, 2021 (MJD $59488.33$). It has right ascension and declination (J2000.0) $01^{\mathrm h}23^{\mathrm m}07.825^{\mathrm s}, -03^\circ11'13.14''$. \cref{fig:field} shows an image of the field before and after the explosion of SN\,2021aaev. Subsequent analysis on the forced photometry indicates that the first detection actually came from the Asteroid Terrestrial-impact Last Alert System (ATLAS; \citealt{ATLAS1}) on MJD 59486.43. The potential host galaxy was identified to be WISEA J012307.75-031111.3 with a redshift of $z=0.1557$. This redshift value is derived from the SN narrow emission features and agrees with the redshift derived from host absorption lines (see \Cref{sec:host}). On Oct 16, 2021, SN\,2021aaev was classified as a SN II \citep{ClassificationTNS} by the Advanced extended Public ESO Spectroscopic Survey of Transient Objects (ePESSTO+; \citealt{Smartt2015}) and then met the $-20$~mag threshold for being a SLSN-II as its peak absolute magnitude reached $<-20$~mag in the optical bands (see \Cref{sec:lightcurve}). Since we also observe persistent narrow H lines in the spectra (see \Cref{sec:spec}), it is spectroscopically a SLSN-IIn.

\subsection{Photometry}

\subsubsection{Optical Photometry from Ground-based Facilities}
\begin{figure*}[t]
    \centering
    \includegraphics[width=\linewidth]{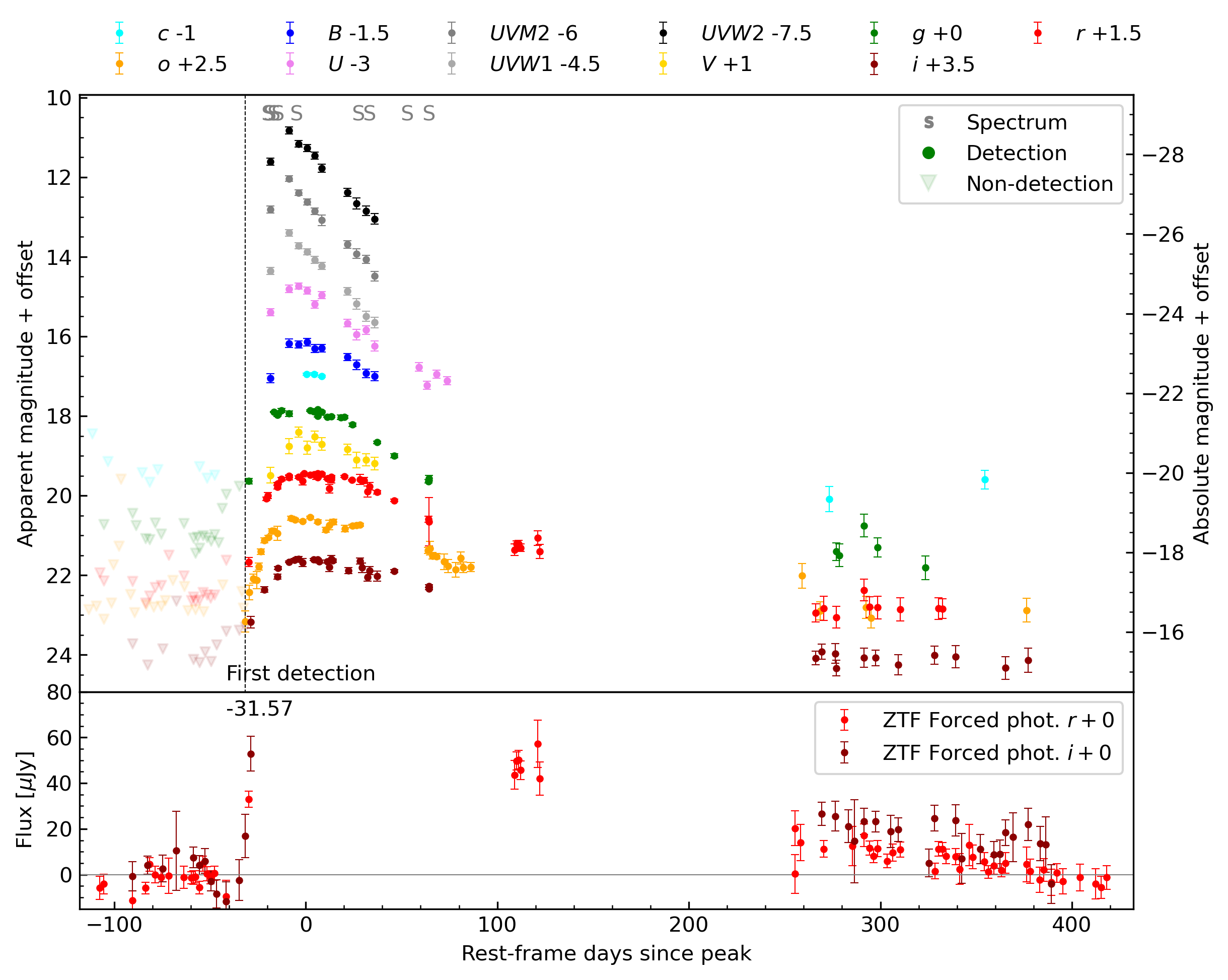}
    \caption{Ultraviolet and optical photometry of SN\,2021aaev. The magnitudes are provided in rest-frame days from peak and are corrected for MW extinction. We use a vertical dashed line to mark the first detection. We mark the epochs of spectral observations with ``S'' symbols at the top of the figure. Bottom panel shows part of the baseline-corrected ZTF $ri$ forced photometry. Beyond $+250$ days, the ZTF $ri$ flux maintained at non-zero flux before going down to zero at about $+400$ days.}
    \label{fig:phot}
\end{figure*}

We obtained multi-band optical photometric observations of SN\,2021aaev from a variety of ground-based facilities. This includes the ZTF, ATLAS \citep{ATLAS1, ATLAS2}, the Spectral Energy Distribution Machine (SEDM; \citealt{SEDM1}) on the Palomar 60-inch telescope, the Katzman Automatic Imaging Telescope (KAIT) at Lick Observatory \citep{Li_2000}, and the Liverpool Telescope (LT) with its IO:O instrument \citep{Liverpool}. Observations were taken in the \textit{gri} bands (ZTF, SEDM, LT), \textit{ri} bands (KAIT), and the \textit{c} and \textit{o} bands (ATLAS). The full photometric dataset spans from Sept 29, 2021 to Nov 22, 2022, or from $-32$ to $+377$ rest-frame days relative to the ATLAS $o$-band peak MJD 59518 (used hereafter; see \Cref{sec:lightcurve}), and is listed in \Cref{tab:photometry}. The Milky Way extinction-corrected photometry (see \Cref{sec:correction}) is shown in \Cref{fig:phot}.

We retrieved the ZTF photometry from the ZTF forced-photometry service \citep{masci2023newforcedphotometryservice}. For each filter, we filtered the data through quality checks, computed the weighted average of the fluxes on a nightly cadence, and converted fluxes to detections or non-detection limits in the AB magnitude system, as described in \cite{masci2023newforcedphotometryservice}.

The remaining photometry was reduced in a similar fashion. The ATLAS photometry was retrieved from the ATLAS forced-photometry service \citep{ATLAS3} and reduced following the pipeline described in \cite{ATLAS2}, which employs Point-Spread Function (PSF)-fitting photometry. Photometry from SEDM and KAIT was processed using the \program{FPipe} pipeline \citep{FPipe}, while the LT photometry reduced using the \program{IO:O} pipeline\footnote{\url{https://telescope.livjm.ac.uk/TelInst/Pipelines/}}.

\subsubsection{Photometry from Swift/UVOT}
We obtained ultraviolet/optical photometry with the 30\,cm space-based Ultraviolet/Optical Telescope (UVOT, \citealt{UVOT}) aboard the Neil Gehrels Swift Observatory \citep{SWIFT} in the \textit{UVW2, UVM2, UVM1, U, B}, and \textit{V} bands, presented in \Cref{fig:phot}. In June 2024, we obtained deep images in all filters to remove the host contamination from the transient photometry. We co-added all sky exposures for a given epoch to boost the S/N using {\tt uvotimsum} in HEAsoft\footnote{\href{https://heasarc.gsfc.nasa.gov/docs/software/heasoft}{https://heasarc.gsfc.nasa.gov/docs/software/heasoft}} version 6.32.2. Afterwards, we measured the brightness of SN with the Swift tool {\tt uvotsource}. The source aperture had a radius of $5''$, while the background region had a significantly larger radius. To remove its contribution from the earlier epochs, we arithmetically subtracted the host flux from the early measurements when the SN was bright. The Swift photometry is listed in \Cref{tab:photometry}.

\subsubsection{Absolute magnitudes} \label{sec:correction}
The absolute magnitude $M$ without host extinction correction in each filter is given by $M=m-\mu-A_{\text{MW},\lambda}-K_{\rm corr}$, where $m$ is the apparent magnitude, $\mu$ is the distance modulus, $A_{\text{MW},\lambda}$ is the Milky Way (MW) extinction at wavelength $\lambda$, and $K_{\rm corr}$ is the cosmological K-correction \citep[see e.g.][]{Kcorrection1,Kcorrection2}. We computed the distance modulus to be $\mu=39.426$~mag, using $z=0.1557$ (see \Cref{sec:host}) and assuming a standard flat-$\Lambda$CDM cosmology with parameters given by \cite{Planck2018}: $H_0=67.66$\,km\,s$^{-1}$~ Mpc$^{-1}$, $\Omega_{M,0}=0.31$. We took $E(B-V)_{\text{MW}}=0.033$~mag\footnote{The MW extinction was calculated with the NASA/IPAC Extragalactic Database's (NED) extinction calculator, which used the \cite{recali} recalibration of the \cite{extinction} infrared-based dust map.} and used the package \program{extinction} and extinction law \texttt{fitzpatrick99} \citep{Fitzpatrick} with $R_{V}=3.1$ to calculate $A_{\text{MW},\lambda}$ for each filter. The K-correction accounts for the difference in the photometric bandpasses in the rest frame and observer frame \citep{hogg2002kcorrection}. We adopted the term $-2.5\log(1+z)$ as an approximation for the full K-correction, as in \cite{Chen_2023}.

\subsection{Spectroscopy}

\begin{figure*}[t]
    \centering
    \includegraphics[width=\linewidth]{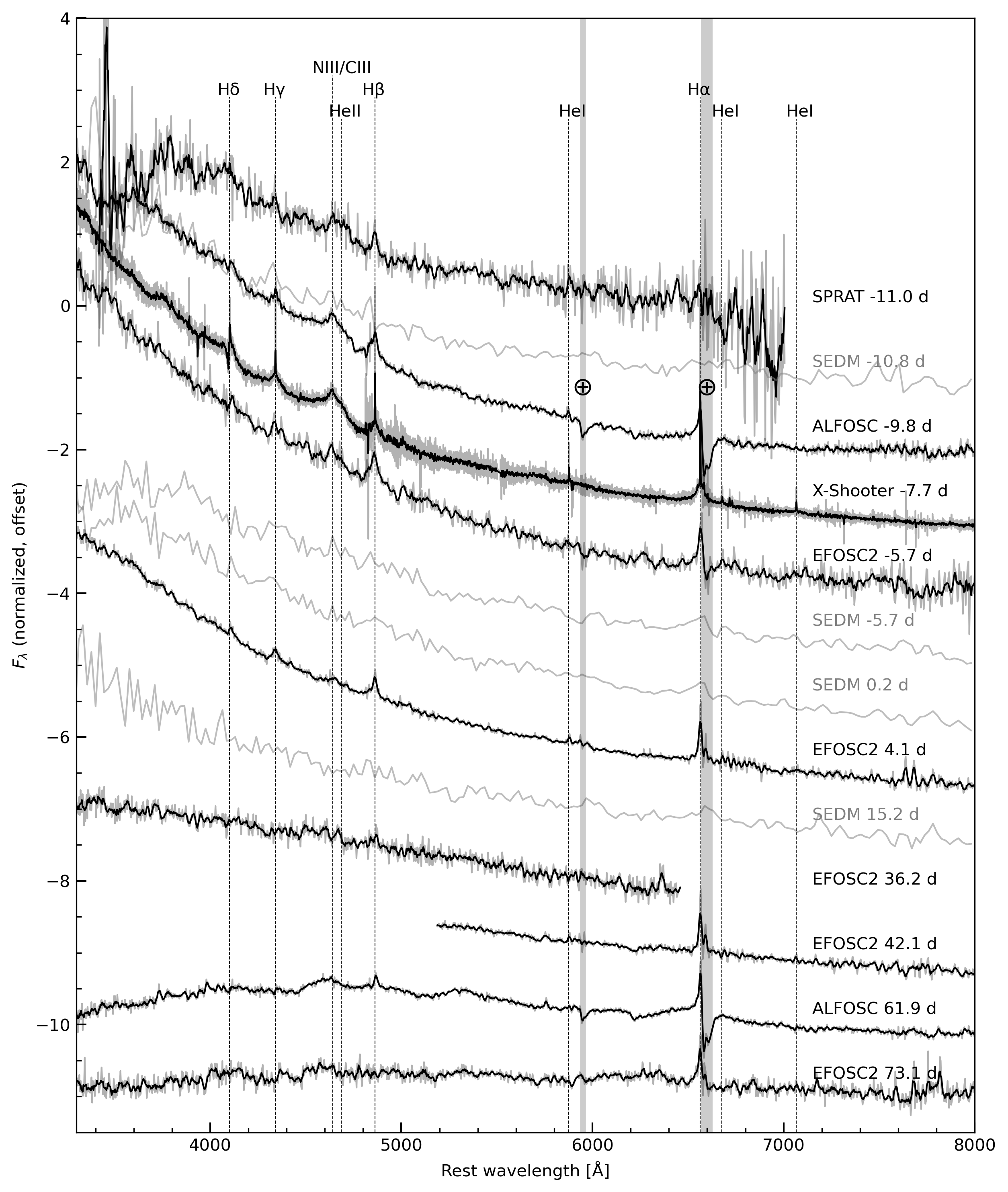}
    \caption{Spectral sequence of SN\,2021aaev from $-20.0$ to $+64.1$ rest-frame days relative to its ATLAS $o$-band peak. We use vertical lines to identify key features. An offset in flux was applied for illustration purposes. The spectra shown in black are smoothed using a Savitzky-Golay filter and the original data are presented in lighter colors.}
    \label{fig:spect}
\end{figure*}

We obtained a total of 13 spectra between Oct\ 10, 2021 and Jan\ 03, 2022, corresponding to rest-frame days $-20.0$ to $+64.1$ relative to the ATLAS $o$-band peak. All spectra were absolutely flux-calibrated using ZTF $gri$ photometry in the following way: we scaled each spectrum according to the average ratio between each spectrum's synthetic $gri$ photometry and the observed ZTF $gri$ photometry at the same epoch. A single scale factor was applied to each spectrum accordingly. The epochs of spectroscopic observations are labeled in \Cref{fig:phot}, and the complete spectral sequence is presented in \Cref{fig:spect}. Information about their phases, exposure times, instruments, grisms used and wavelength range can be found in \Cref{tab:spectroscopy}.

The first spectrum, obtained during the rise of the SN, was taken with the SPRAT spectrograph \citep{SPRAT} on the 2m LT. We reduced and flux calibrated this spectrum using a custom \program{Python} pipeline, utilizing the packages \program{Matplotlib} \citep{Hunter_2007_Matplotlib}, \program{NumPy} \citep{van_der_Walt_2011_Numpy}, \program{SciPy} \citep{Virtanen_2020_SciPy} and \program{Astropy} \citep{2013A&A...558A..33A}. We also used the L.A.Cosmic algorithm \citep{vanDokkum2001a} to mask cosmic rays. For flux calibration we used the standard star Hiltner 102 \citep{Stone_1977_StandardStar}, which was observed on the same night. Airmass differences between science and standard star observations were then corrected for using Table 1 from La Palma Technical Note No. 31\footnote{\url{https://www.ing.iac.es/Astronomy/observing/manuals/ps/tech_notes/tn031.pdf}}. Four additional low-resolution spectra were obtained with SEDM \citep{Kim_2022_SEDM}, automatically reduced by \program{pysedm} \citep{pysedm}.

Two spectra were obtained with ALFOSC on the 2.56m Nordic Optical Telescope (NOT). One was taken during the rise using grism 4 and a 1" wide slit, while the other was taken post-peak with grism 4 and a 1.3" wide slit, depending on the observing conditions. Reduction was performed using the \program{PyNOT} pipeline\footnote{\url{https://github.com/jkrogager/PyNOT}} in a standard fashion. 

We obtained five spectra with the ESO Faint Object Spectrograph and Camera 2 (EFOSC2) on the New Technology Telescope (NTT) at La Silla Observatory as part of the ePESSTO+ program \citep{Smartt2015, ePESSTO+}. Observations were made with a 1.0" wide slit and grisms 11, 13, and 16, and we reduced the data using the \program{PESSTO} pipeline\footnote{\url{https://github.com/svalenti/pessto}}. We attempted to correct for telluric absorption with the pipeline, with varying degrees of success. We also stitched spectra from the same day taken with different grisms by averaging overlapping regions.

A higher-resolution X-Shooter spectrum was obtained with ESO's 8.2\,m Very Large Telescope (VLT; \citealt{X-shooter}) at Paranal Observatory. Observations were performed in nodding mode with 1.0", 0.9", and 0.9" wide slits for the UV, visible (VIS), and near-infrared (NIR) arms, respectively. The data were reduced following \cite{Selsing2019a}. In brief, we first removed cosmic rays with the tool \program{astroscrappy}\footnote{\href{https://github.com/astropy/astroscrappy}{https://github.com/astropy/astroscrappy}}, which is based on the cosmic-ray removal algorithm by \cite{vanDokkum2001a}. Afterwards, the data were processed with the X-shooter pipeline v3.3.6 and the ESO workflow engine \program{ESOReflex} \citep{XShooter1,XShooter2}. The UVB and VIS-arm data were reduced in stare mode to boost the S/N. The individual rectified, wavelength- and flux-calibrated two-dimensional spectra files were co-added using tools developed by J. Selsing\footnote{\href{https://github.com/jselsing/XSGRB_reduction_scripts}{https://github.com/jselsing/XSGRB\_reduction\_scripts}}. The NIR data were reduced in nodding mode to ensure a good sky-line subtraction. In the third step, we extracted the one-dimensional spectra of each arm in a statistically optimal way using tools by J. Selsing. During this step, we also correct for the strongest telluric absorption features of the coadded VIS-arm spectrum with the software package \program{MOLECFIT} version 3.3.6 \citep{Molecfit1,Molecfit2}. Finally, the wavelength calibration of all spectra was corrected for barycentric motion. The spectra of the individual arms were stitched by averaging the overlap regions.

\section{Photometric Analysis} \label{sec:phot}

\subsection{Rest-frame multi-band light curves}\label{sec:lightcurve}
\begin{figure}[t]
    \centering
    \includegraphics[width=\linewidth]{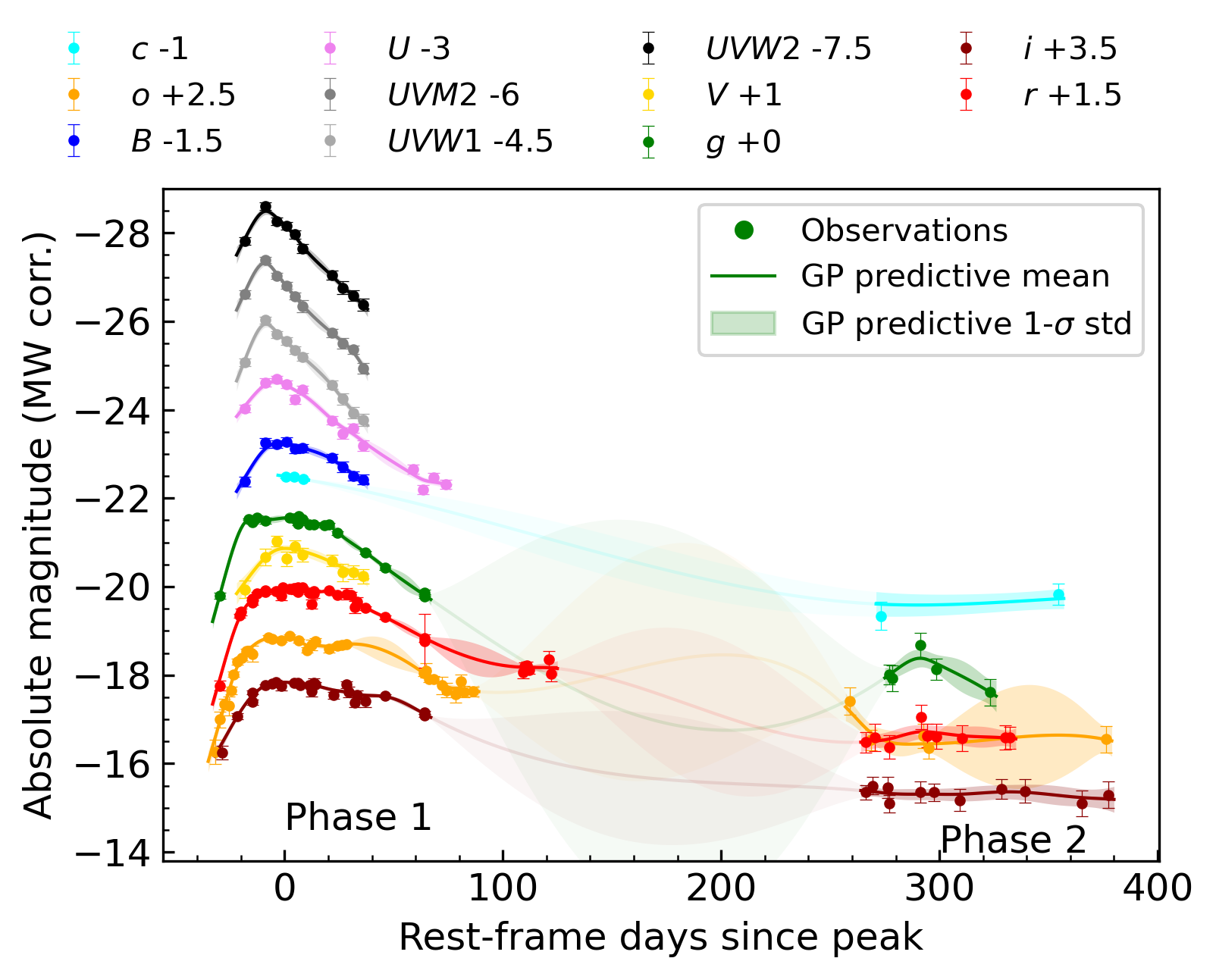}
    \caption{Interpolated multi-band light curves of SN\,2021aaev using GP. We interpolated each band separately by a single mean function with a 1$\sigma$ spread. The  predictive variance is large at the solar conjunction phase between phase 1 and phase 2, and we avoided using interpolation in this gap.}
    \label{fig:GP}
\end{figure}

To characterize the different phases of SN\,2021aaev, we converted the time axis to the object's rest-frame. The rest-frame time is defined as the time elapsed since the peak of the ATLAS $o$-band light curve of SN\,2021aaev (see \Cref{fig:GP}). We chose the ATLAS $o$ band because it is the most densely sampled band in the rising and peaking phases.

\begin{figure}[t]
    \centering
    \includegraphics[width=\linewidth]{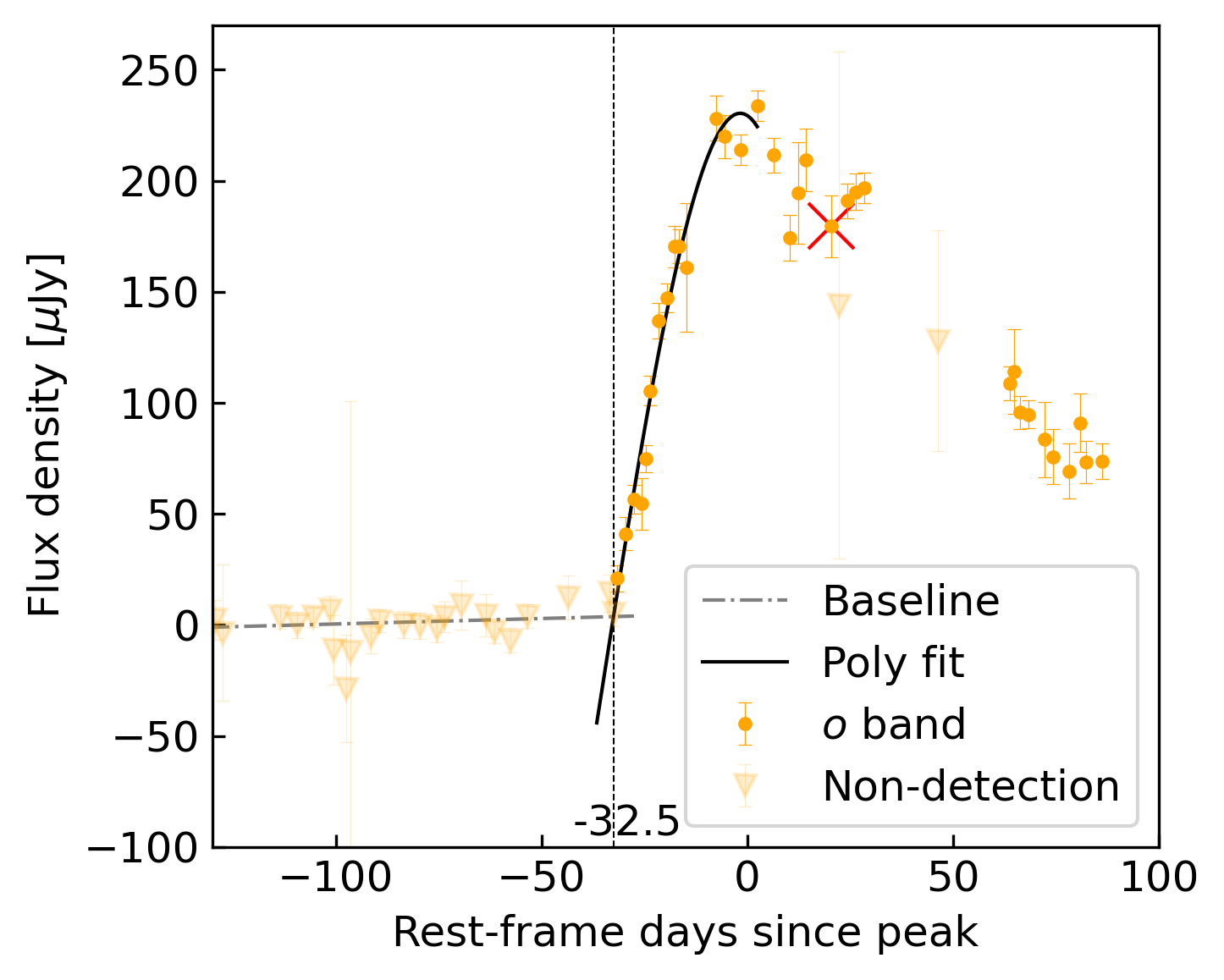}
    \caption{Polynomial fit using the $o$-band photometry to find the day of first light. We fitted all detections until the peak with a 3$^{\text{rd}}$ order polynomial. The baseline of non-detections was fitted with a straight line. The red cross marks the turning point of a potential bump.}
    \label{fig:poly}
\end{figure}

To estimate the peak time of the light curve, we need a smooth light curve and this was done by data interpolation. A popular method to perform data interpolation is Gaussian process (GP) regression, a method in Bayesian inference to model priors as a probability distribution over functions. For a comprehensive guide, see \cite{GP}. We ran a GP regression on each band separately for this multi-band dataset\footnote{Some recent works assigned a correlation between different bands and GP-interpolated all bands in a single run by assigning an effective wavelength to each band. However, little is understood about the color evolution of SLSNe-II, and we do not want to correlate the time scale of the light curve in one band to another. Hence, we did not adapt such a technique.}. The regression was performed using \program{george} \citep{2015ascl.soft11015F} with a Mat\'ern $5/2$ kernel, and the result is shown in \Cref{fig:GP}. To estimate the day of first light, we fitted a third order polynomial to the $o$-band data from the first detection to peak, and solved for its intersection with a baseline from the non-detections (see \Cref{fig:poly}).

Now we can characterize the multi-band light curves of SN\,2021aaev. In the rest frame, we estimate the first light to be $-32.5\pm1.0$ days before the $o$-band peak. In the ATLAS $o$ band, the light curve shows an $1/e$ rise time of $25\pm1$ days, with the peak occurring on MJD $59518\pm1$ at an absolute magnitude $-21.35\pm0.03$. The peaks in the optical $groi$ bands are smooth. The $o$ band then declined on a time scale of $73\pm1$ days to $1/e$ of its maximum. Some potential bumps may be identified in the $o$ and $r$ bands at $\sim$ day 40--50, but there is no further observational evidence, apart from the three epochs of rise in $o$ band after the red cross marked in \Cref{fig:poly}. In the UV, the $UVW2$ light curve reached the peak at MJD $59509\pm1$, which is nine days earlier than the optical peak. It has a much sharper peak, followed by a faster decline of $21\pm1$ days to $1/e$ of its maximum. No secondary bumps can be identified in the UV bands. A summary of the characterization in the $r,o, UVM2$ bands is presented in \Cref{tab:light_curve}. After solar conjunction, some epochs in the optical were observed with ZTF and ATLAS, labeled as ``phase 2'' in \Cref{fig:GP}. An apparent bump was observed in the $g$ band, possibly due to sparse data and large measurement uncertainties, but no obvious fluctuations or color change were seen in the $roi$ light curves. 

\begin{table}[t]
    \centering 
    \begin{tabular}{c c c c}
        \hline\hline
        Parameter & $r$ & $o$ & $UVM2$\\
        \hline
        Peak MJD [day] & $59519^{+1}_{-1}$ & $59518^{+1}_{-1}$ & $59509^{+1}_{-1}$ \\
        Peak mag [mag] & $-21.46\pm0.01$ & $-21.35\pm0.03$ &$-21.32\pm0.07$\\
        $t_{\text{rise},1/e}$ [day] & $26\pm1$ & $25\pm1$ & -\\
        $t_{\text{decline},1/e}$ [day]& $62\pm1$ & $73\pm1$ & $21\pm1$\\
        $t_{\text{rise},10\%}$ [day]& $32\pm1$ & $31\pm1$ & -\\
        $t_{\text{decline},10\%}$ [day] & $162\pm1$ & $146\pm1$ & $47\pm1$\\
        \hline
    \end{tabular}
    \caption{Characterization of key parameters of the $r, o, UVM2$ band light curves for SN\,2021aaev. Other than peak magnitude, all other parameters are estimated in flux space. We conservatively estimate the uncertainty of any time parameters to be $\pm1$ day because the interpolation was done on every integer MJD. The $t_{\text{rise/decline},1/e}$ or $t_{\text{rise/decline},10\%}$ is defined as the time from/to $1/e$ or $10\%$ of the maximum flux to/from the maximum flux on the rise/decline. We were unable to compute the rise timescale for the $UVM2$ band because only one point was observed before the peak.}\label{tab:light_curve}
\end{table}

\subsection{Light-curve comparison}
\begin{figure}[t]
    \centering
    \includegraphics[width=\linewidth]{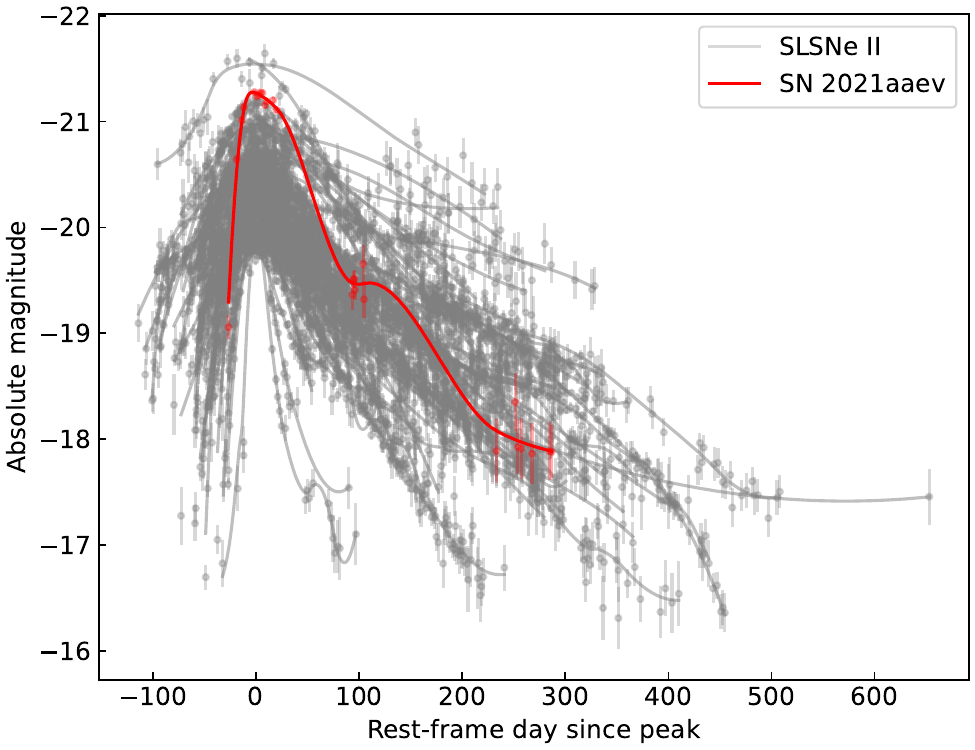}
    \caption{Comparison between the ZTF $r$-band light curve of SN\,2021aaev with the rest of the ZTF sample in \cite{pessi2025}. All interpolations are done using automated loess regression (ALR; \citealt{ALR}).}
    \label{fig:sample_rband}
\end{figure}

\begin{figure}[t]
    \centering
    \includegraphics[width=\linewidth]{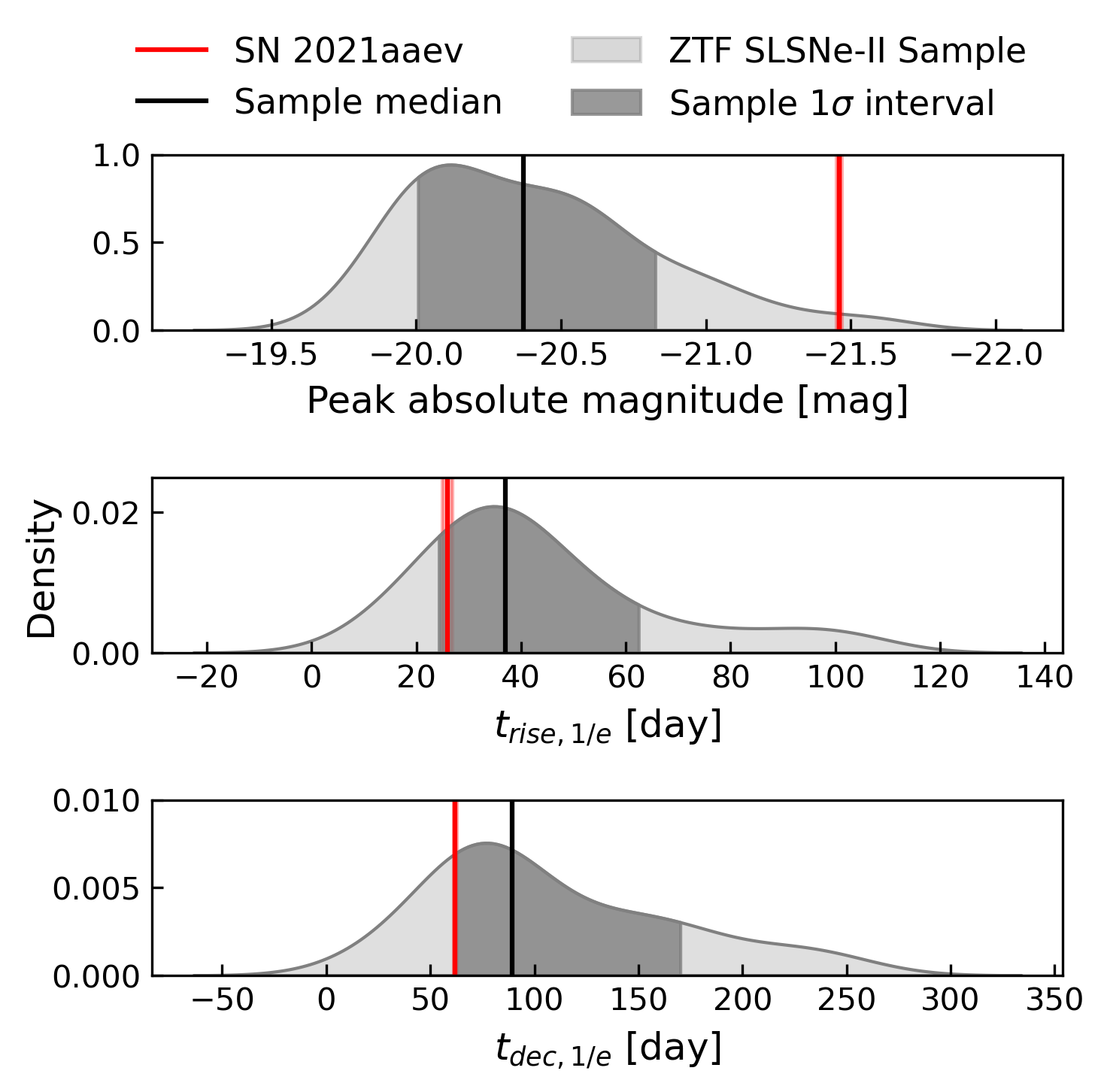}
    \caption{Kernel density estimate of the $r$-band parameters of the ZTF sample in \cite{pessi2025}, with the black vertical line indicating the sample mean and SN\,2021aaev marked by the red vertical line.}
    \label{fig:sample}
\end{figure}

We compare the $r$-band light curve of SN\,2021aaev with the largest SLSN-II light-curve sample to date in \cite{pessi2025}\footnote{A slightly different set of cosmological parameters were used in \cite{pessi2025}, but this will not significantly affect the inference on light curve parameters.}, and show this comparison in \Cref{fig:sample_rband}. SN\,2021aaev's $r$-band light curve is distinguished by its higher absolute magnitude but shows non-exceptional temporal behavior. We also plot kernel density estimates of the $r$-band peak absolute magnitude as well as the $1/e$ rise and decline times in \Cref{fig:sample}. SN\,2021aaev appears luminous compared to the median, falling outside the 1$\sigma$ interval for absolute magnitude. In terms of timescales it is not exceptional. Its slightly faster rise and decline rates remain within the 1$\sigma$ range of the sample. 

\begin{figure}[t]
    \centering
    \includegraphics[width=\linewidth]{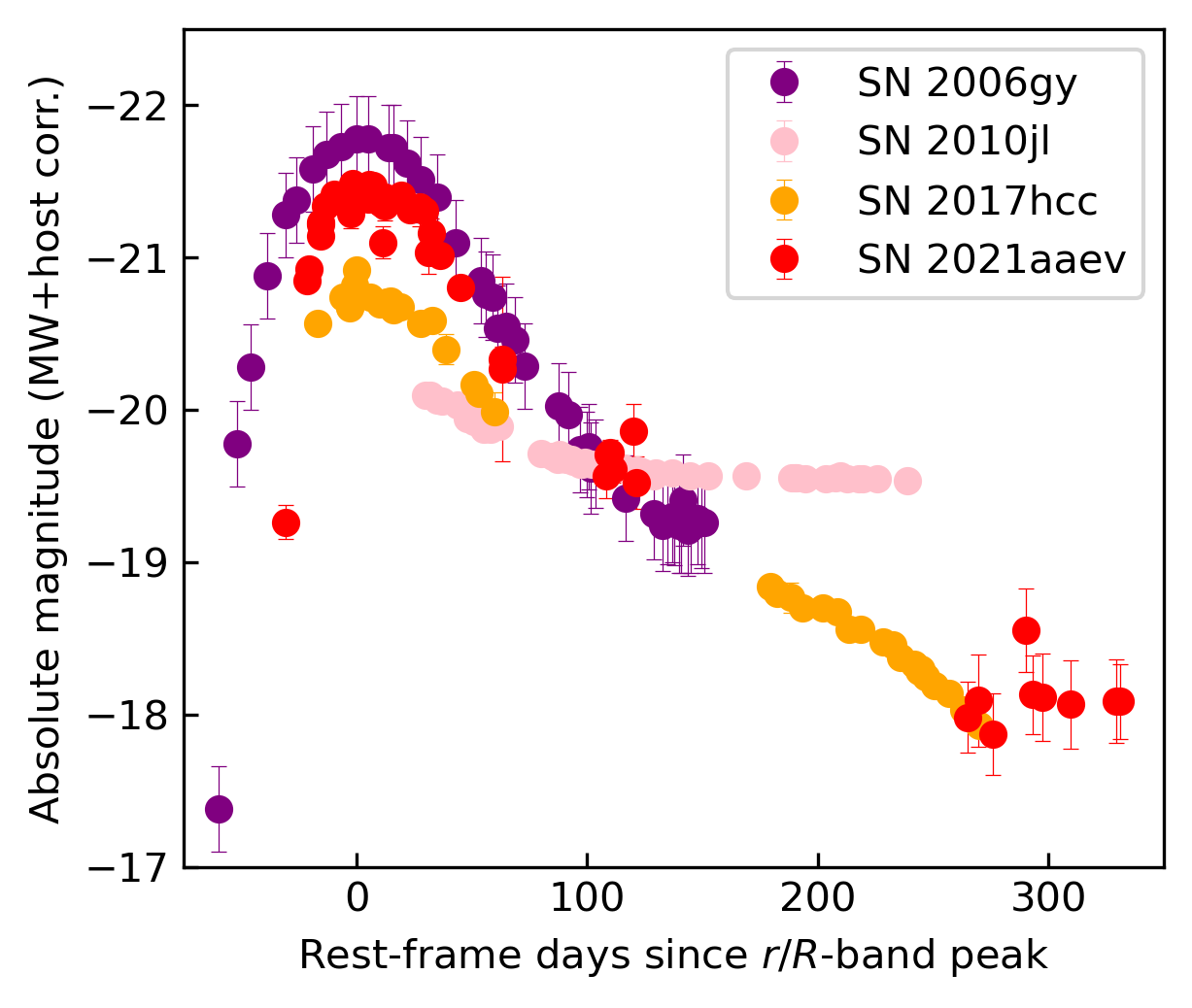}
    \caption{Comparison between the $r$-band light curve of SN\,2021aaev with SN\,2017hcc \citep{Moran_2023}, SN\,2010jl \citep[e.g.][]{Fransson_2014} and SN\,2006gy \citep[$R$ band, see e.g.][]{2007ApJ...666.1116S}. The photometry for SN\,2010jl and 2006gy are corrected for host extinction, while SN\,2017hcc and 2021aaev has negligible or undetermined host extinction.}
    \label{fig:rband_comparison}
\end{figure}

We also compare the $r$-band light curve of SN\,2021aaev with the $r$/$R$-band light curves of other well-studied H-rich, luminous, CSM-interacting SNe, plotted for SN\,2006gy, SN\,2010jl and SN\,2017hcc in \Cref{fig:rband_comparison}. The $r$ peak of SN\,2021aaev is brighter than SN\,2017hcc and perhaps SN\,2010jl but fainter than SN\,2006gy. In terms of timescales, SN\,2021aaev has a shorter rise time compared to SN\,2017hcc and SN\,2006gy (both about 50--60 days). For the decline timescale, SN\,2021aaev has a similar post-maximum decline to SN\,2017hcc and SN\,2006gy but they all decline faster than SN\,2010jl. This is consistent with the result from the comparison with the ZTF SLSN-II sample. All four objects are long-lasting, with SN\,2021aaev and SN\,2017hcc at $-18$~mag even at 300 days after peak and SN\,2010jl at $-18$~mag at 250 days after discovery. For the three well-studied SNe, massive and/or extensive CSM were invoked to explain the photometry; we expect a similar powering mechanism and CSM properties for SN\,2021aaev. CSM-powering models are explored in \Cref{sec:model}.

\subsection{Pseudo-bolometric light curve, Temperature and Radius}
\begin{figure}[t]
    \centering
    \includegraphics[width=\linewidth]{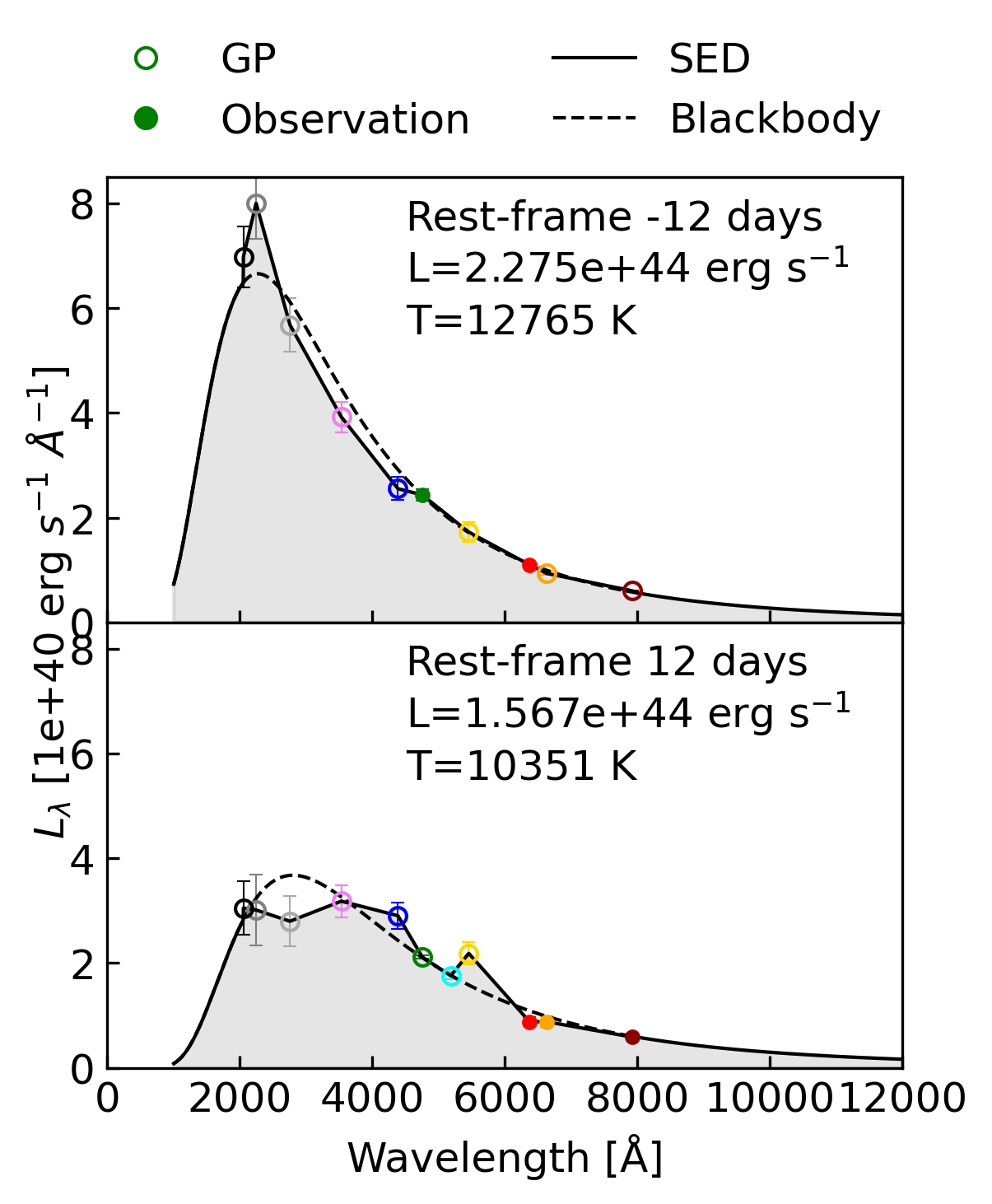}
    \caption{The SED of SN\,2021aaev, using one epoch before peak ($-12$ days) and one epoch after peak ($+12$ days) as examples. The total luminosities are corrected with a far-UV (from 1000 \AA) and a NIR (to 24,000\AA) blackbody. The colors for the filters are the same as those in \Cref{fig:GP}.}
    \label{fig:SED}
\end{figure}

\begin{figure}[t]
    \centering
    \includegraphics[width=\linewidth]{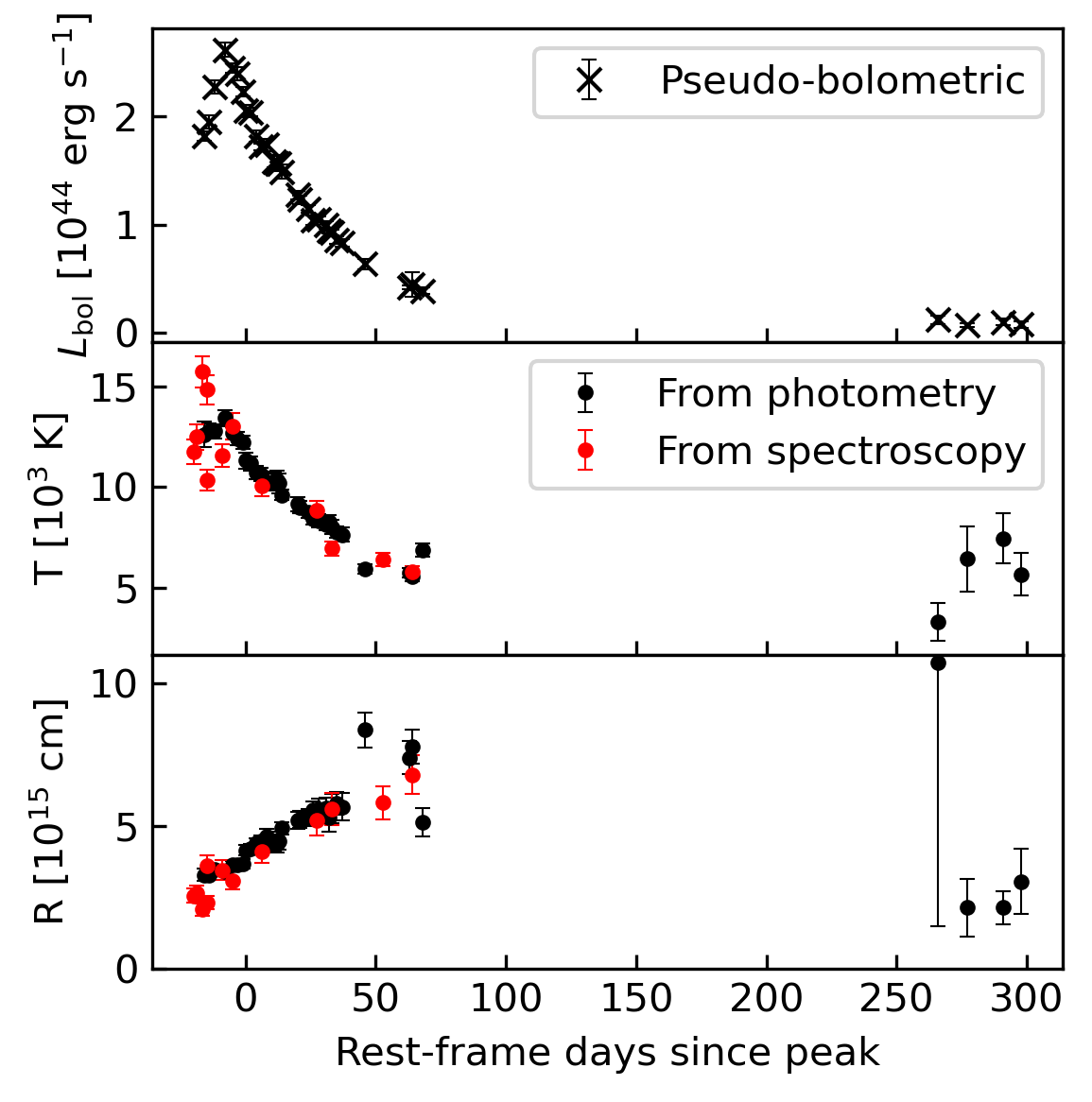}
    \caption{\textit{Top}: UV-optical (2000--8000 \AA) pseudo-bolometric light curve of SN\,2021aaev, corrected using black-body fit to the far-UV (from 1000 \AA) and the NIR (to 24000 \AA). \textit{Middle} \& \textit{bottom}: photospheric blackbody temperature and radius evolution of SN\,2021aaev from its multi-band photometry and absolute-calibrated spectra (see \Cref{sec:spec}.).}
    \label{fig:t_and_r_evolution}
\end{figure}

We constructed the pseudo-bolometric light curve of SN\,2021aaev from its UV and optical light curves in the following way. We selected epochs with observations in at least 2 bands, and in total there are 34 such epochs. For these 34 epochs, we used the observed bands and the GP value of unobserved bands to fit black body curves. These bands covered about $\lambda=2000$--$8000$ \AA. For extrapolations, for the NIR we attached a blackbody tail until $\lambda=24000$ \AA{} and for the far-UV end we attached a blackbody curve until $\lambda=1000$ \AA. To get the luminosity we connected points with straight lines between $\lambda=2000$--$8000$ \AA{} and integrated the area under the solid lines (see \Cref{fig:SED}). The UV-optical blackbody-corrected pseudo-bolometric light curve from this construction is shown at the top of \Cref{fig:t_and_r_evolution}.

The pseudo-bolometric light curve of SN\,2021aaev peaks at $2.62\times10^{44}$~erg~s$^{-1}$, 9 days before the ATLAS $o$ band, and the total energy released in radiation is estimated to be $1.4\times10^{51}$~erg. This level of radiative output is in principle attainable via standard core-collapse explosion mechanisms such as a delayed neutrino-driven explosion alone \citep{CCSNe_mechanisms}. However, achieving such a high radiative efficiency would require a substantial fraction of the ejecta's kinetic energy to be converted into radiation. In the presence of a dense CSM, this conversion can occur efficiently through shock interaction, which serves as a mechanism to tap into the kinetic energy reservoir of the ejecta. From a simple kinetic perspective, efficient energy conversion requires a strong deceleration of the ejecta which favors a ``heavy'' CSM scenario, where the mass of CSM is comparable or even exceeds that of the SN ejecta \citep[see e.g.][]{Ofek2014, khatami2024}. This is discussed in \Cref{sec:model}.

When fitting the blackbody we also obtained the evolution of the estimated photospheric blackbody temperature and radius, as shown in \Cref{fig:t_and_r_evolution}. These estimates help trace the thermal and geometric evolution of the emitting region over time. The temperature and radius derived from spectroscopy (presented in \Cref{fig:spect}) are also included in the plot and are generally consistent with those derived from photometry. The blackbody temperature follows a similar evolutionary trend as the pseudo-bolometric light curve, peaking near the UV maximum (rest-frame $-9$ days) at approximately $15{,}000$~K and declining to around $5000$~K by +60 days. The high temperature during the early phase suggests a substantial UV contribution, which is characteristic of SLSNe. This is consistent with the flash-ionization features observed (see \Cref{sec:flash}) and is a sign of intense interaction with CSM. Meanwhile, the blackbody photospheric radius expands until approximately rest-frame $+60$ days. During this phase, the photosphere is located in the unshocked CSM and continues to expand as more of the CSM becomes ionized and optically thick due to the ongoing shock interaction (e.g., \citealt{Chevalier1994}), reflecting the expansion of the ionized region. After this point, the photosphere recedes in radius. This decline in radius is a consequence of decreasing optical depth as the outer layers thin out and the emission originates from progressively inner regions, within the shocked CSM or the reverse-shocked SN ejecta.

\subsection{Color evolution}
\begin{figure}[t]
    \centering
    \includegraphics[width=\linewidth]{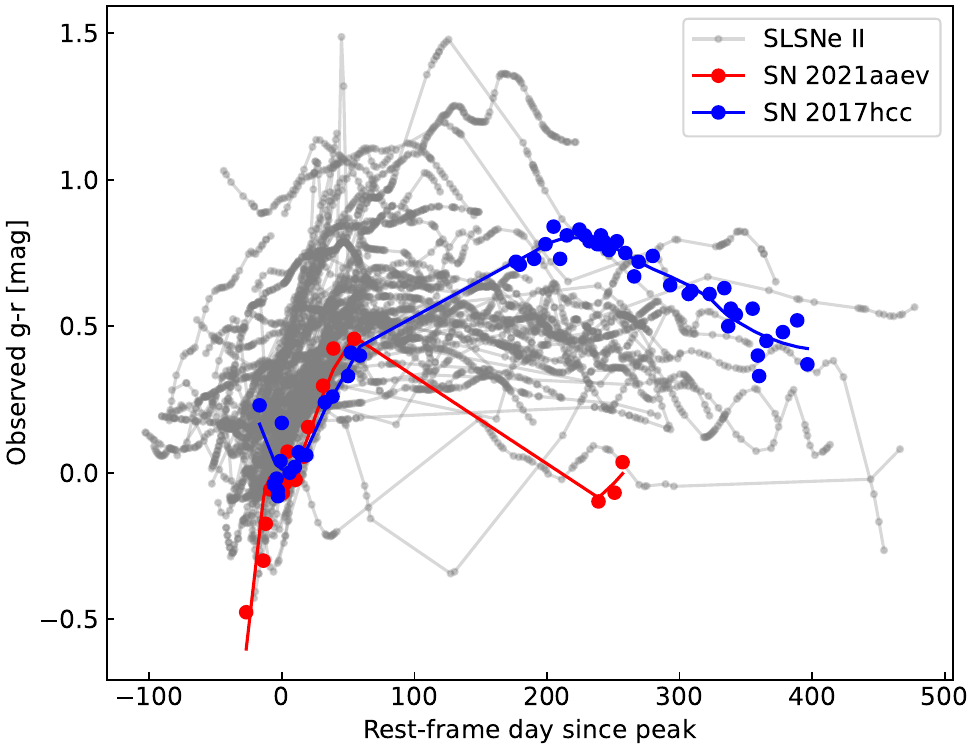}
    \caption{The ZTF $g-r$ color evolution of SN\,2021aaev, SN\,2017hcc \citep{Moran_2023} and the SLSN-II sample in \cite{pessi2025}. All interpolations are done using ALR \citep{ALR}.}
    \label{fig:color}
\end{figure}

We calculated the $g-r$ color of SN\,2021aaev, and plot the color evolution in \Cref{fig:color}. For comparison, we include SN\,2017hcc and SLSNe-II from the ZTF sample in \cite{pessi2025}. SN\,2021aaev evolves from a relatively blue $g-r$ colour of approximately $-0.5$~mag at $\sim -30$ days to a redder color of $0.5$~mag by $\sim80$ days. At very late times ($>250$ days), the color declines again, reaching around $-0.1$~mag.

Compared to other SLSNe-II in the ZTF sample, SN\,2021aaev exhibits a relatively blue early photospheric phase, suggesting a higher blackbody temperature ($15\,000$~K). Despite this, its early-time color evolution is broadly consistent with the rest of the ZTF sample. Notably, at late times ($\gtrsim 200$ days), SN\,2021aaev becomes significantly bluer while most SLSNe-II, such as SN\,2017hcc, maintain a redder plateau. This reversion to bluer colors may reflect renewed heating of the photosphere, possibly driven by ongoing or renewed CSM interaction which appears to occur on a faster timescale than in most other SLSNe-II. This blue-ward shift was also observed in SN\,2017hcc at a later phase ($+230$--$460$ days) and was explained as the appearance of a forest of \ion{Fe}{2} lines which affects the broadband photometry \citep{Moran_2023}. Alternatively, such a trend is also seen in the radioactive $^{56}$Ni-powered tail phase of some SNe II \citep[see e.g.][]{Yang_2024}. However, in the case of SLSNe-II, the high luminosity at these late phases would require an implausibly large $^{56}$Ni mass if radioactive decay were the dominant energy source. A third explanation is that the late-time photometry has a significant contribution from a light echo (see e.g. \citealt{2006gy_3}). This can be tested by comparing the peak SED, scaled by a scattering law of the form $\lambda^{-n}$ (where $0 \leq n \leq 2$), to the late-time SED. However, the prominent $i$-band flux at $\gtrsim300$ days (see \Cref{fig:phot}) is difficult to reproduce with such a scattering law, suggesting that a light echo is unlikely to be the dominant cause of the observed bluening. Thus, in the absence of late-time spectra, CSM interaction remains the more likely explanation.

\section{Spectroscopic Analysis} \label{sec:spec}

\subsection{Spectral Evolution}

\begin{figure*}[t]
    \centering
    \includegraphics[width=\linewidth]{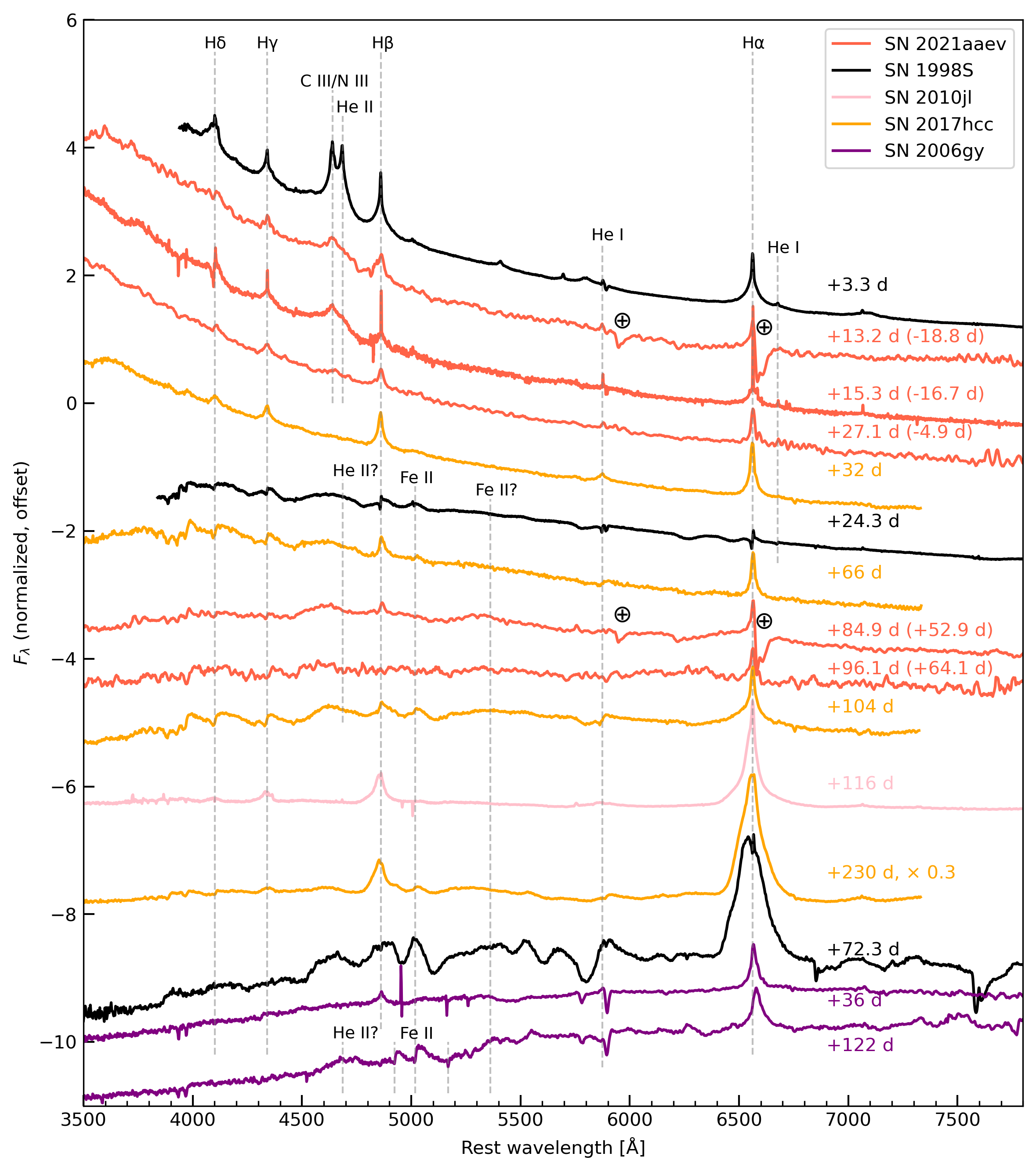}
    \caption{Spectral evolution of SN\,2021aaev compared with H-rich, superluminous SN\,2006gy, SN\,2010jl, and SN\,2017hcc, and a SN IIn flasher SN\,1998S. All fluxes are normalized with respect to the continuum near H$\alpha$. The phases without brackets represent days since first light of each SN, while the phases with bracket indicate rest-frame days relative to the maximum light for SN\,2021aaev.}
    \label{fig:evolution}
\end{figure*}

The full spectral sequence of SN\,2021aaev is presented in \Cref{fig:spect}. A selection of spectra are presented and compared to spectra of well-studied SNe IIn and SLSNe-IIn in \Cref{fig:evolution}. A strong telluric line falls onto the red side of the H$\alpha$ of SN\,2021aaev. This affects the reliability of the red-side of the H$\alpha$ profile in most spectra, although successful telluric correction has been applied to the $+15.37$ days X-Shooter spectrum. 

In the early photospheric phases (up to $+27.1$ days), the spectra were characterized by a blue continuum and Lorentzian-winged narrow Balmer lines. The blue continuum comes from the hot, optically thick, expanding photosphere located in the ionized CSM. No ejecta features were observed. This spectral morphology is common among SNe IIn in their early photospheric phase, where narrow emission lines arise from slow-moving CSM beyond the photosphere and are broadened into Lorentzian wings via electron scattering (see \Cref{sec:escatter} for details). During this phase, there was also an evolving feature around $4600$ \AA{} that we identify as from flash-ionization (possibly a blend of \ion{C}{3} $\lambda$4649, \ion{N}{3} $\lambda\lambda$4634, 4641, and \ion{He}{2} $\lambda$4686). However, the flash features in SN\,2021aaev shows no doubly-peaked narrow components, whereas many other flash spectra such as the $+3.3$ days spectrum of SN 1998S do. This is discussed in \Cref{sec:flash}. 

As the spectra evolved near maximum-light, the flash feature began to fade and was hardly visible by $+27.1$ days. At a comparable phase, the first spectrum of SN\,2017hcc was taken ($+32$ days) which was strikingly similar to the $+27.1$ days spectrum of SN\,2021aaev, and no flash features could be identified. SN\,2006gy, on the other hand, has a much redder early-time spectrum ($+36$ days) compared to its photometrically similar counterparts. Following this phase, a broad feature, likely photospheric He, appeared at about $4500$--$4600$ \AA{}. This broad feature has been observed in the $+66$ days to $+104$ days spectra of SN\,2017hcc, and is also present in some SNe IIn, albeit at a much earlier phase (e.g. $+24.3$ days in SN\,1998S). A forest of Fe lines blending into a broad feature appears between 5000 and 5500 \AA, which is also seen in SN\,1998S in its $+24.3$ days spectrum and the $+122$ days spectrum of SN\,2006gy. 

The H$\alpha$ line in the $+84.9$ and $+96.1$ days spectra remains narrow and strong, with no evidence of broadening. This behavior contrasts with that of many less-luminous SNe IIn in their late photospheric phase, where broad features from the ejecta begin to emerge (e.g. $+72.3$ days in SN\,1998S), though exceptions exist (e.g. SN 1994W, \citealt{1994W3, 1994W1, 1994W2}; SN 2009kn, \citealt{2009kn}). At the $+84.9$ days phase, the spectral evolution of SN\,2021aaev shows striking similarity with SN\,2017hcc. The persistence of narrow Balmer lines in both cases indicates continued CSM interaction at these late phases and suggests that the ejecta have been strongly or entirely decelerated, implying $M_{\text{CSM}} \ge M_{\text{ej}}$.

From $+96.1$ days onward, we have no spectra of SN\,2021aaev. For SN\,2017hcc, the $+230$ days spectrum showed a relatively broader H$\alpha$ ($1690\pm20$ km s$^{-1}$, \citealt{Moran_2023}) with a blueshifted peak. Similar broadening and a blue excess is seen in the $+116$ days spectrum of SN\,2010jl \citep{Fransson_2014}.

\subsection{Flash-ionization features} \label{sec:flash}

We observe blended flash-ionization features of \ion{He}{2} $\lambda$4686 and possibly \ion{C}{3} $\lambda$4649/\ion{N}{3} $\lambda\lambda$4634, 4641 in the early photospheric phase of SN\,2021aaev. These highly-ionized lines are strong evidence for the presence of dense and confined CSM \citep{Gal_Yam_2014, Khazov_2016}, which typically last only a few days during the rise phase \citep{Yaron_2017,Bruch_2023}. Flash features are commonly observed in H-rich SNe, with estimates suggesting that at least 30\% of all H-rich SNe display such signatures if early spectra were obtained \citep{Bruch_2023}. SN\,2021aaev is the first SLSNe-IIn where flash spectroscopy have been observed, and another SLSN-II (with broad Balmer lines) with flash features is SN\,2023gpw (estimated peak $r$-band magnitude $=-21.47$~mag; \citealt{Kangas_2025}). 

\begin{figure}[t]
    \centering
    \includegraphics[width=\linewidth]{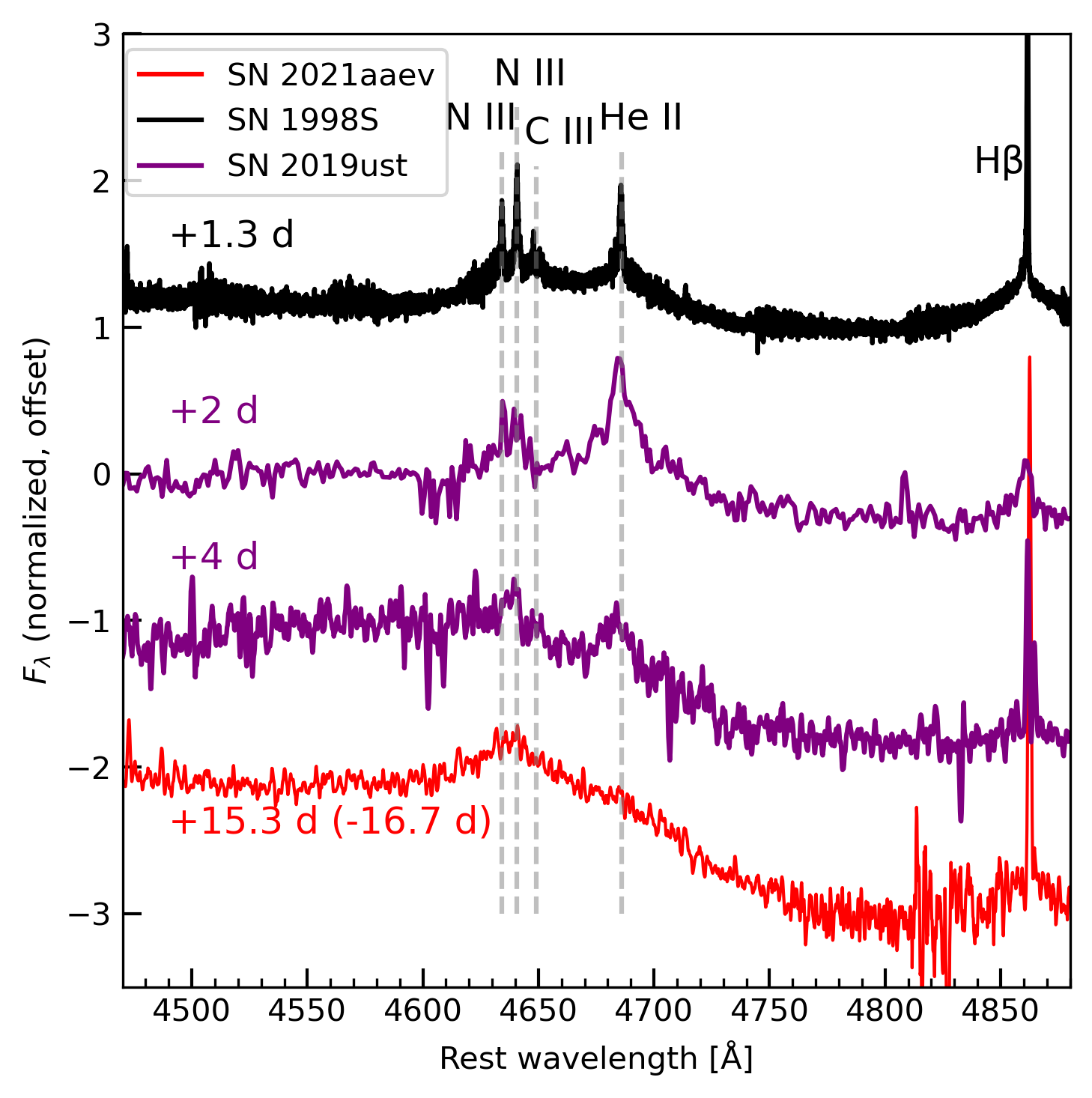}
    \caption{Evolution of the \ion{He}{2} $\lambda4686$ and \ion{C}{3} $\lambda$4649/\ion{N}{3} $\lambda\lambda$4634, 4641 flash features of SN\,2021aaev, as compared to that of SN\,1998S and SN\,2019ust.}
    \label{fig:flash_evolution}
\end{figure}

While the \ion{He}{2} $\lambda$4686 flash feature in many other flashers exhibit distinct narrow components, no such components are observed in SN\,2021aaev. In addition, the \ion{He}{2} $\lambda4686$ feature in SN\,2021aaev appears consistently weaker than the nearby feature at $\sim4645$ \AA{} in all observed flash spectra, whereas in many other flashers (e.g. see the first spectrum of SN\,1998S in \Cref{fig:evolution}), the \ion{He}{2} line is comparably strong, if not stronger. This may indicate that the flash features caught in SN\,2021aaev were at a later phase ($\sim 15$ days since first light) compared to most other flashers. For instance, the evolution of SN\,2019ust \citep{Bruch_2023} in \Cref{fig:flash_evolution} shows that this flash feature is fast-evolving and broadens and weakens as time progresses. Broad flash features are sometimes referred to as ``ledge'' features \citep[see e.g.][]{Soumagnac_2020,Pearson_2023} and are explained as either blueshifted \ion{He}{2} $\lambda4686$ \citep{Gal-Yam_2011} or a blend of other highly-ionized C, N, O lines \citep{Dessart_2017}. We adopted the second explanation for SN\,2021aaev because no other high-velocity features are observed in any spectra at a similar phase. 

\begin{figure}[h] 
\centering 
\includegraphics[width=\linewidth]{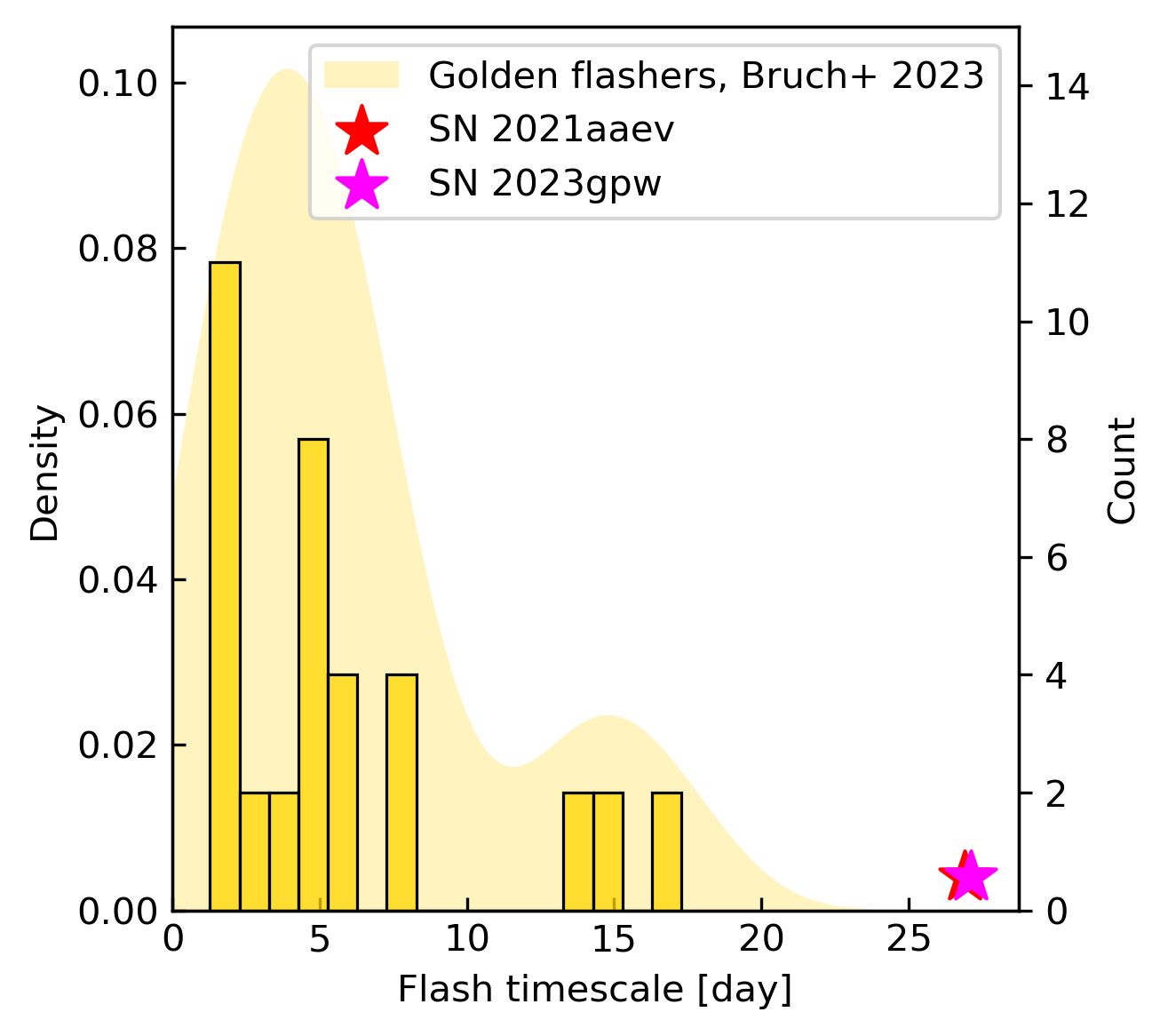} 
\caption{Kernel density estimate and histogram of flash timescale of the golden flasher sample \citep{Bruch_2023} and SN\,2021aaev.} 
\label{fig:flash} 
\end{figure}

\begin{figure*}[t]
    \centering
    \includegraphics[width=\linewidth]{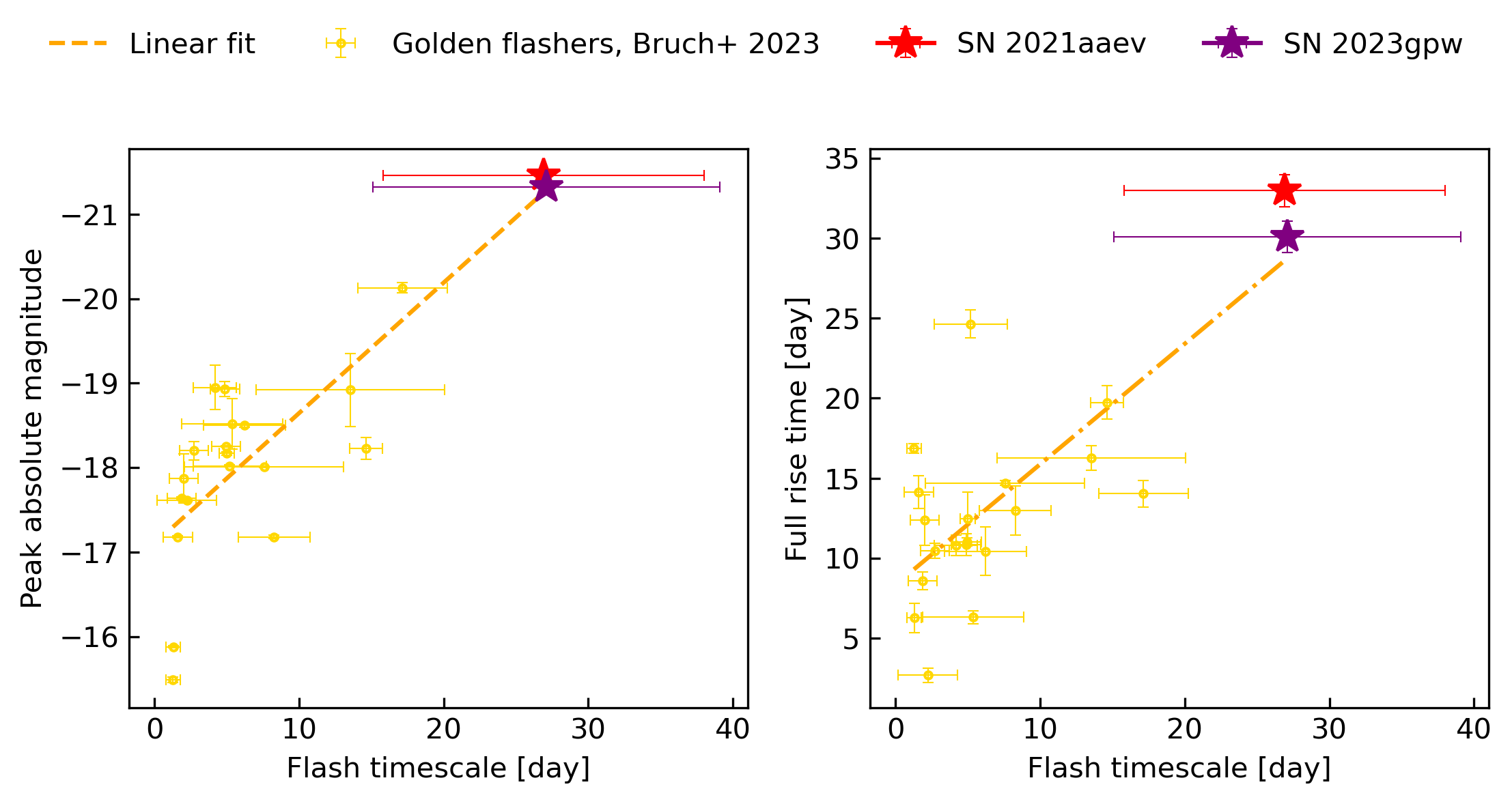}
    \caption{\textit{Left}: flash timescale against peak absolute magnitude in $r$ band. \textit{Right}: flash timescale against rise time in $r$ band. The plot builds on figure 16 in \cite{Bruch_2023} with newly added SN\,2021aaev and SN\,2023gpw \citep{Kangas_2025}.}
    \label{fig:flash_comparison}
\end{figure*}

The long flash-timescale observed in SN\,2021aaev is also noteworthy. \cite{Bruch_2023} compiled a ``golden flasher'' sample of H-rich SNe, in which they defined the flash timescale as the duration from first light to the midpoint between the last flash spectrum and the first non-flash spectrum. However, this definition may not be appropriate for superluminous SN\,2021aaev and SN\,2023gpw, both of which lack early spectral coverage. In the case of SN\,2021aaev, there is an 11-day gap between first light and the earliest flash spectrum obtained with SPRAT, and a similar situation applies to SN\,2023gpw \citep{Kangas_2025}. To improve this estimate, we redefine the start of the flash timescale for the two SLSNe-II as the midpoint between first light and the first flash spectrum, rather than using first light directly. This revised estimate still places SN\,2021aaev and SN\,2023gpw (both about 27 days) among the longest flashers, as shown in \Cref{fig:flash}.

\cite{Bruch_2023} found correlations between the flash timescale, peak luminosity, and rise time. We added SN\,2021aaev and SN\,2023gpw to their correlation plots (see \Cref{fig:flash_comparison}) and recalculated the relationships. The two SLSN-II flashers have much longer rise times and brighter peaks than most objects in the golden flasher sample, but they still follow the trend. For peak magnitude versus flash timescale, we obtained a Pearson correlation coefficient of $-0.835$ ($p=0.000003$). For rise time versus flash timescale, we obtained a Pearson correlation coefficient of $0.783$ ($p=0.0003$). Both indicate strong correlations, which may reflect underlying dependencies on the mass and radial extent of the CSM. However, the sample remains too small to draw definitive conclusions, in particular for the SLSN-II class.

\subsection{Electron scattering} \label{sec:escatter}

Quantitative information on the optical depth of flash features in CSM can be extracted by modeling their Lorentzian wing broadening as the result of electron scattering. Photons emitted from the forward shock front scatter off free electrons in the CSM, leading to broadened line profiles. A Monte Carlo code, \program{escatter}\footnote{\url{https://github.com/eScatter/e-scatter}}, has been developed to model H$\alpha$ line profiles under the assumption that the CSM follows a single power-law density profile, $\rho = \rho_0 r^{-s}$. It has been applied to SLSNe-II such as SN\,2021adxl \citep{Brennan_2024}.

\begin{figure*}[t]
    \centering
    \includegraphics[width=\linewidth]{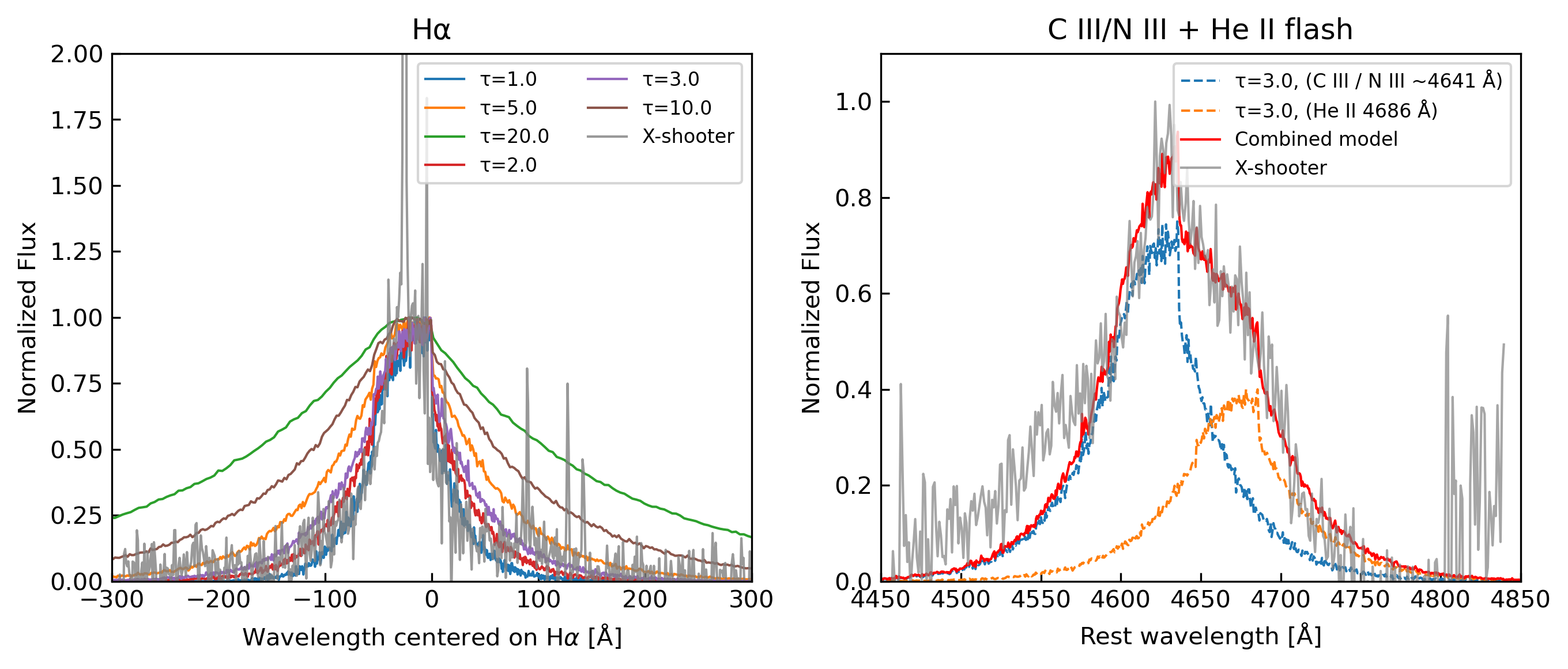}
    \caption{Comparison between observed Lorenetizan-winged features in the X-Shooter spectrum of SN\,2021aaev and models generated by \program{escatter}. The code allows one to change wind velocity $v_{\text{w}}$, shock velocity $v_{\text{s}}$ and CSM density exponent $s$, but they do not impact the line profile. We fixed them as: $v_{\text{w}}=100$ km s$^{-1}$, $v_{\text{s}}=2500$ km s$^{-1}$, $s=2$. \textit{Left}: Observed H$\alpha$ and a grid of line profile models with different Thomson optical depths $\tau$. \textit{Right}: Observed blended flash features at 4600--4700 \AA{} and two arbitrarily scaled line profile (with fixed separation) models with $\tau=3$.}
    \label{fig:escatter}
\end{figure*}

We assume that for SN\,2021aaev, the broadening seen in the flash features and H$\alpha$ (with a velocity dispersion of $1399\pm38$\,km\,s$^{-1}$) in the $-16.7$ days X-Shooter spectrum arise from the same electron scattering process. For H$\alpha$, we ignored the narrow component and compare the Lorentzian wing with a grid of \program{escatter} models with Thomson optical depth ranging from 1 to 20 (left of \Cref{fig:escatter}). Models with lower optical depth ($\tau=1,2$) agreed the best with the observed H$\alpha$ profile. For the blended flash feature, we assumed that the broadening has a contribution from the \ion{He}{2} $\lambda$4686 line and a second contribution from an unidentified line (possibly \ion{C}{3} $\lambda$4649 or \ion{N}{3} $\lambda$4634, averaged to 4641). We compared the observation with a \program{escatter} model of $\tau=3$ (right of \Cref{fig:escatter}), and arbitrarily scale the two model lines' amplitudes but keeping their separation fixed. A blue deficit exists in the model but it otherwise agrees well with the observation. 

From the analysis above we conclude that the two Lorentzian-winged features seen in the X-Shooter spectrum of SN\,2021aaev came from regions with slightly different radial optical depths. The blended flash features likely arise from a somewhat deeper layer, in the ionized, unshocked CSM just outside the forward shock, where the UV flash from shock breakout ionizes the material, and electron scattering shapes the line profiles. The CSM must be dense enough to produce scattering wings, but not so dense as to obscure the flash completely, consistent with a moderately optically thick, extended CSM environment. 

\section{Light Curve Modeling} \label{sec:model}
Following all the analysis on different pieces of information from photometry and spectroscopy above, we now attempt to depict a self-consistent picture of SN\,2021aaev. We explore possible light-curve models that explain the powering of SN\,2021aaev and are consistent with the conclusions from the spectral analysis.

\subsection{Analytical CSM-interaction models}
\begin{figure}[h]
    \centering
    \includegraphics[width=\linewidth]{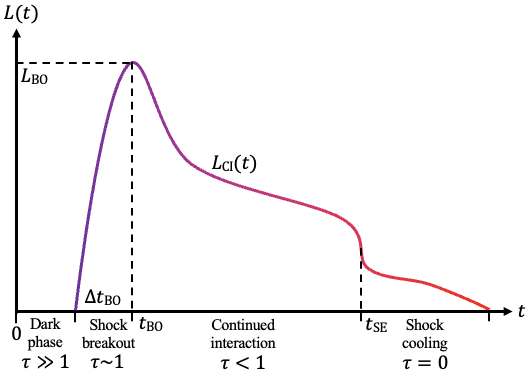}
    \caption{Schematic diagram of different phases of a CSM interaction-powered SN light curve, adapted from \cite{khatami2024} for the interior breakout scenario. Initially the shock front is in an optically thick dark phase. As the shock propagates radially outward it transitions into an optically thin region and we could witness shock breakout inside the CSM with sharply rising luminosity. Then, the continued CSM-ejecta interaction powers the light curve. At $t_{\text{SE}}$, we would see the light curve plummet as the shock emerges from the CSM edge.}
    \label{fig:Khatami_Kasen_paper}
\end{figure}

We searched for generic analytical models that are applicable to an extensive CSM interaction-powered SLSNe. \cite{khatami2024} introduced an analytical framework to fit light curves of CSM interaction-powered SNe, under the assumptions that the inner CSM edge is much smaller that the outer edge and the ejecta reach homologous expansion before colliding with the CSM. In this framework,  the ``shock breakout'' is defined as the moment when the forward shock front reaches a low enough optical depth for photons to escape, while the moment that the shock exits the CSM entirely is referred to as ``shock emergence''. Two key parameters are introduced: mass ratio $\eta=M_{\text{CSM}}/M_{\text{ej}}$ and ``breakout'' parameter $\xi$ which depends on $\eta$, as well as the outer radius of CSM $R_{\text{CSM}}$ and the initial ejecta velocity $v_{\text{ej}}$. These define four light-curve regimes: heavy ($\eta > 1$) or light ($\eta < 1$) CSM, and breakout at the edge ($\xi > 1$) or interior ($\xi < 1$).

To interpret the long time scale and sustained luminosity of SN\,2021aaev, we adopt the interior shock breakout regime (schematic diagram shown in \Cref{fig:Khatami_Kasen_paper}), which is consistent with the spectroscopic evidence for prolonged CSM interaction. We fitted the interior-breakout analytical scaling relations in \citet[][sect.~3]{khatami2024} with $\kappa_e=0.34$\,cm$^2$\,g$^{-1}$ to the bolometric light curve, and corrected using the factors in \citet[][fig.~17]{khatami2024} from numerical simulations. Since we did not observe a shock emergence phase of SN\,2021aaev corresponding to $t_{\text{SE}}$ in \Cref{fig:Khatami_Kasen_paper}, we set the estimated shock emergence earlier limit to the last observed photometry epoch and assigned a large uncertainty of 200 days. The model involves 7 free parameters: $M_{\text{CSM}}$, $M_{\text{ej}}$, $R_{\text{CSM}}$, $v_{\text{ej}}$, the CSM density exponent $s$, the outer ejecta density exponent $n$, and a breakout exponent $k_0$ where $k_0=0$ refers to edge breakout and $k_0=1$ is immediate breakout once the interaction begins. However, there are degeneracies between parameters, especially between $M_{\mathrm{ej}}$ and $v_{\mathrm{ej}}$. Due to the absence of broad ejecta features, such as P-Cygni profiles or nebular-phase emission lines, we cannot directly infer the ejecta velocity from observations. Therefore, to improve inference, we fixed $n=10$ and $v_{\text{ej}}=11000$\,km\,s$^{-1}$, which is typical for CCSNe. We applied uniform priors: $M_{\text{CSM}}\in[0,25]\,M_{\odot}$, $M_{\text{ej}}\in[0,25]\,M_{\odot}$, $R_{\text{CSM}}\in[1\times10^{14},5\times10^{16}]$\,cm, $k_0\in[0.59,1]$ and $s\in[0,3]$.

\begin{figure*}[t]
    \centering
    \includegraphics[width=0.49\textwidth]{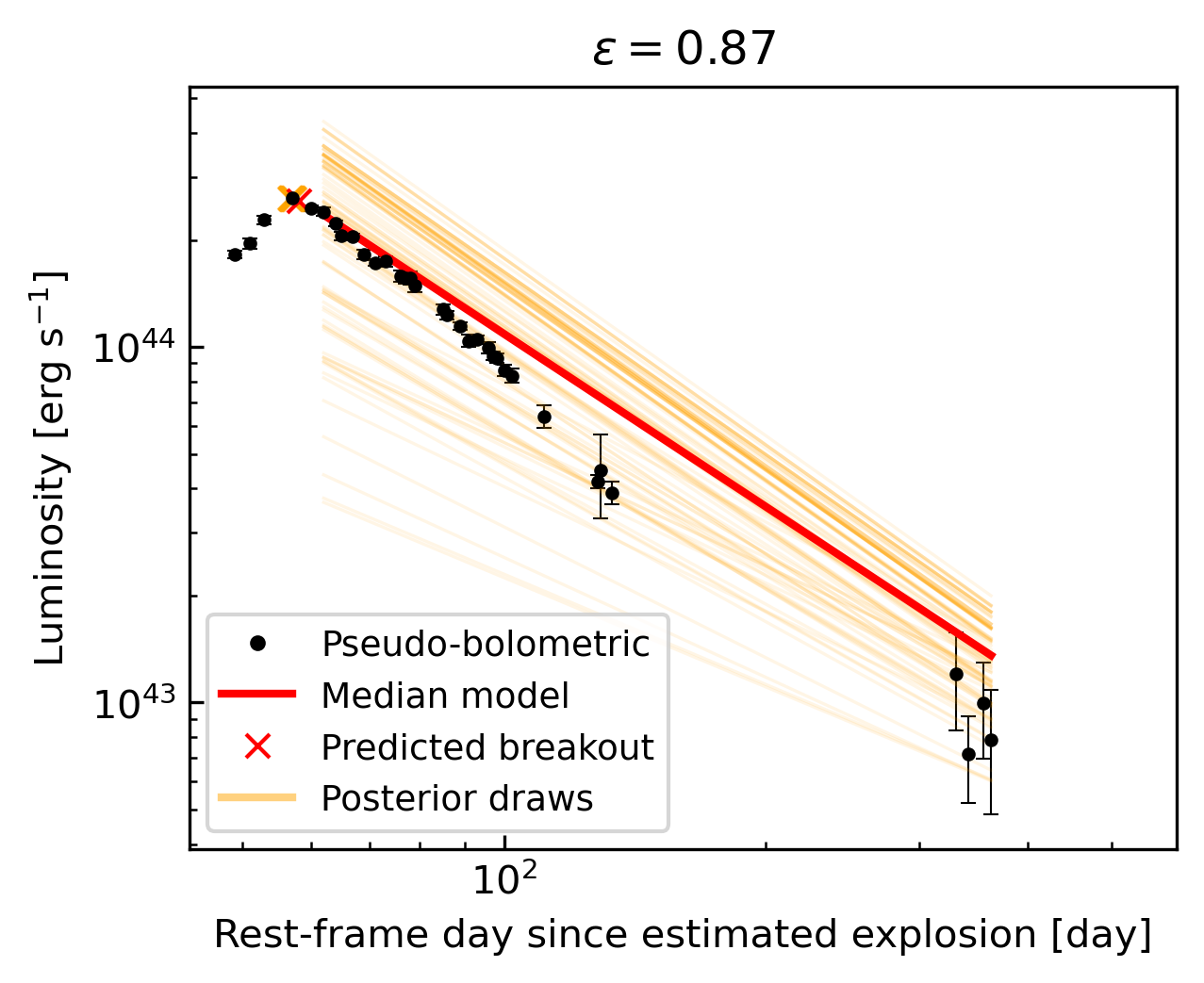}
    \includegraphics[width=0.49\textwidth]{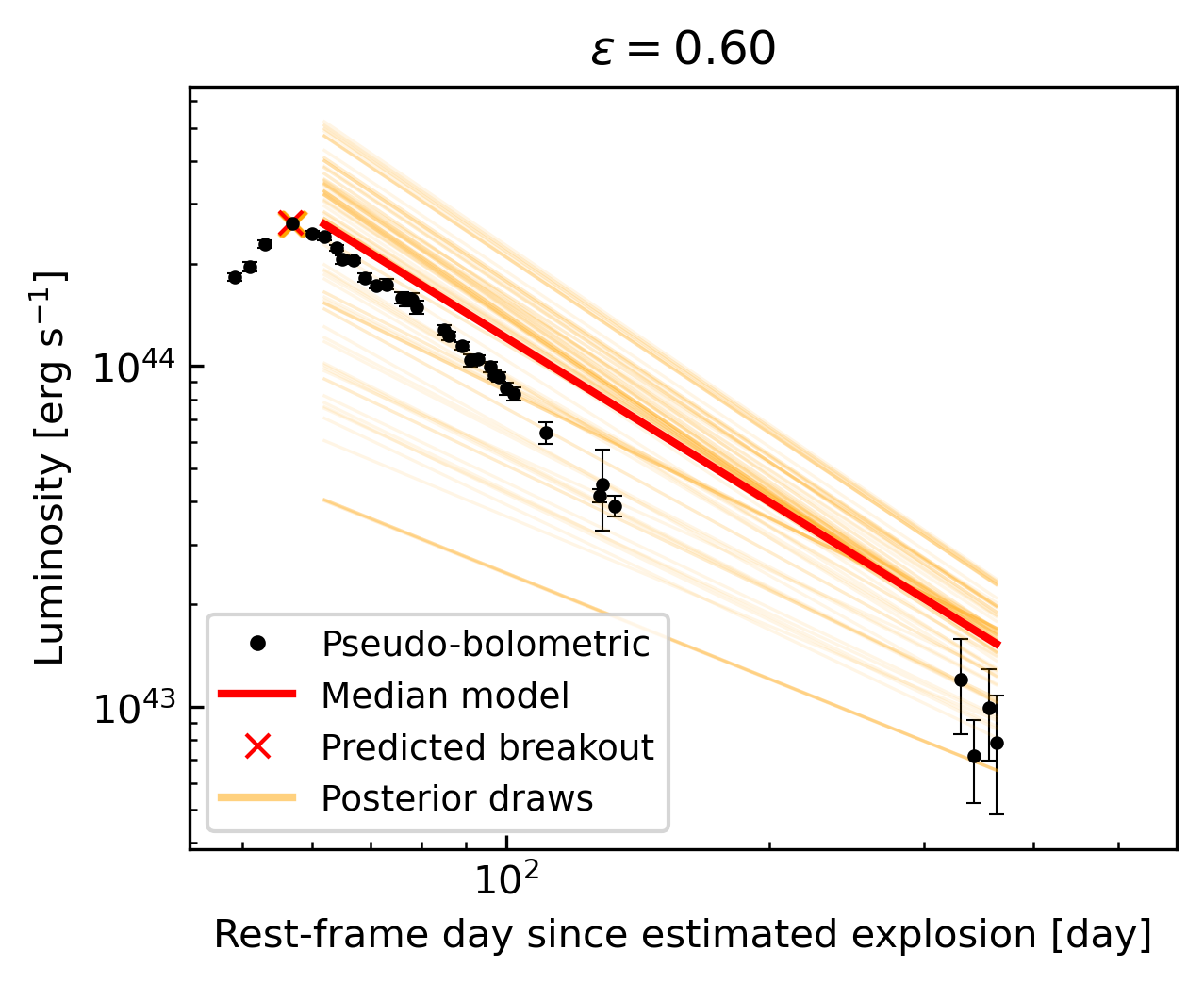}
    \caption{
    \textit{Left}: Comparison between the bolometric light curve constructed from observed photometry and analytical models in \cite{khatami2024}, with no constraint on energy conversion efficiency $\epsilon$. The best-fit model gives $\epsilon=0.87$. \textit{Right}: The same model with additional constraint that $\epsilon=0.60$. We draw the plots in log-log scale for better visualization.    
    }
    \label{fig:Khatami_Kasen_fit}
\end{figure*}

We found that the best-fit (median) scenario is for an ejecta of $M_{\text{ej}}=1.34^{+0.06}_{-0.06}\,M_{\odot}$ running into a CSM of $M_{\text{CSM}}=12.9^{+3.8}_{-3.9}\,M_{\odot}$ and $R_{\text{CSM}}=1.57^{+0.35}_{-0.29}\times10^{16}\text{cm}$, shown in the left of \Cref{fig:Khatami_Kasen_fit} (corner plot shown in \Cref{fig:Khatami_Kasen_parameters}). The result is also summarized in the first row of \Cref{tab:models_parameter}. The fitted bolometric light curve matches the breakout peak and captures the overall trend of decline well (with a slight overshoot). This clearly supports the heavy-interior regime, where the SN ejecta ran into a more massive CSM, continuously interacting with the CSM and depositing most of the kinetic energy into radiation (with an efficiency calculated to be $\epsilon=0.87$). The existence of a continued interaction phase has spectral evidence support up to $+96.6$ days since first light, as we see narrow H$\alpha$ and H$\beta$ lines with decreasing strength during the decline phase of the light curve. This physical picture is also consistent with the electron scattering analysis in \Cref{sec:escatter}. The shock breakout phase corresponds to the epochs where we see flash features and the optical depth transitions from $\tau\gg1$ to $\tau\sim1$. Beyond this phase, $\tau<1$ and extensive wind-like CSM density may not be large enough to support an observable flash, so no flash features should be observed afterwards.

\begin{table*}[t]
    \centering
    \begin{tabular}{l c c c c c c c c c}
    \hline\hline
    Model & $\epsilon$ & $M_{\text{ej}}$ & $v_{\text{ej}}$ & $M_{\text{CSM}}$ & $R_{\text{CSM}}$ & $k_0$ & $s$ & $B_p$ & $P_0$\\
     & & [$M_{\odot}$] & [$10^3$ km s$^{-1}$] & [$M_{\odot}$] & [$10^{16}$ cm] & & & [$10^{14}$ G] & [ms]\\
    \hline
    CSM$^*$ & $0.87$ & $1.34^{+0.06}_{-0.06}$ & $11$ & $12.9^{+3.8}_{-3.9}$ & $1.57^{+0.35}_{-0.29}$ & $0.85^{+0.10}_{-0.11}$ &$1.55^{+0.98}_{-1.02}$ &  - & - \\
    CSM$^{**}$ & $0.60$ & $1.96^{+0.09}_{-0.08}$ & $11$ & $14.4^{+4.4}_{-4.1}$ & $1.78^{+0.36}_{-0.31}$ & $0.85^{+0.10}_{-0.10}$ & $1.51^{+1.03}_{-1.04}$ &  - & - \\
    Magnetar & - & $1.17^{+0.05}_{-0.05}$  & $9.85^{+0.13}_{-0.14}$ & - & - & - & - & $1.53^{+0.96}_{-0.45}$ & $3.95^{+0.56}_{-0.56}$\\
    \hline
    \end{tabular}
    \caption{Best-fit parameters of various models for SN\,2021aaev. CSM*: the ``heavy-interior'' analytical CSM-interaction model in \cite{khatami2024}. CSM**: the same CSM-interaction model with additional constraint that energy conversion efficiency $\epsilon=0.60$. Magnetar: magnetar-powering model in \citep{Nicholl2017}, implemented using \program{Redback} \citep{Sarin2024}.}
    \label{tab:models_parameter}
\end{table*}

Nevertheless, the simple analytical model presented above has many limitations. Firstly, the derived kinetic energy of $E_{\text{KE}}=1.62\times10^{51}$~ergs would require an exceptionally high energy conversion efficiency of 0.87. If we impose the additional constraint that the conversion efficiency $\epsilon=0.6$ (close to that of SN\,2006gy inferred in \citealt{khatami2024}) by scaling the numerical factors simultaneously in the analytical model, then the median model becomes an ejecta of $M_{\text{ej}}=1.96^{+0.09}_{-0.08}\,M_{\odot}$ running into a CSM of $M_{\text{CSM}}=14.4^{+4.4}_{-4.1}\,M_{\odot}$ and $R_{\text{CSM}}=1.78^{+0.36}_{-0.31}\times10^{16}\text{cm}$, see the right of \Cref{fig:Khatami_Kasen_fit} (corner plot shown in  \Cref{fig:Khatami_Kasen_parameters2}). This does not change the ``heavy-interior'' conclusion, but the fit is in general poorer. Secondly, the constraint on $s$ is weak, meaning that we cannot use this model to accurately estimate the density profile of the CSM. Another limitation is that it does not provide any quantitative prediction on the rising part of the light curve. A more detailed modelling of the rising light curve and CSM density profile, e.g. similar to the approach presented in \cite{Cosentino2025} for low-massive CSM, could be pursued in future work to improve our understanding of the shock-CSM interaction in SN\,2021aaev.

\subsection{Magnetar Model} \label{sec:magnetar_model} 
\begin{figure}[t]
    \centering
    \includegraphics[width=\linewidth]{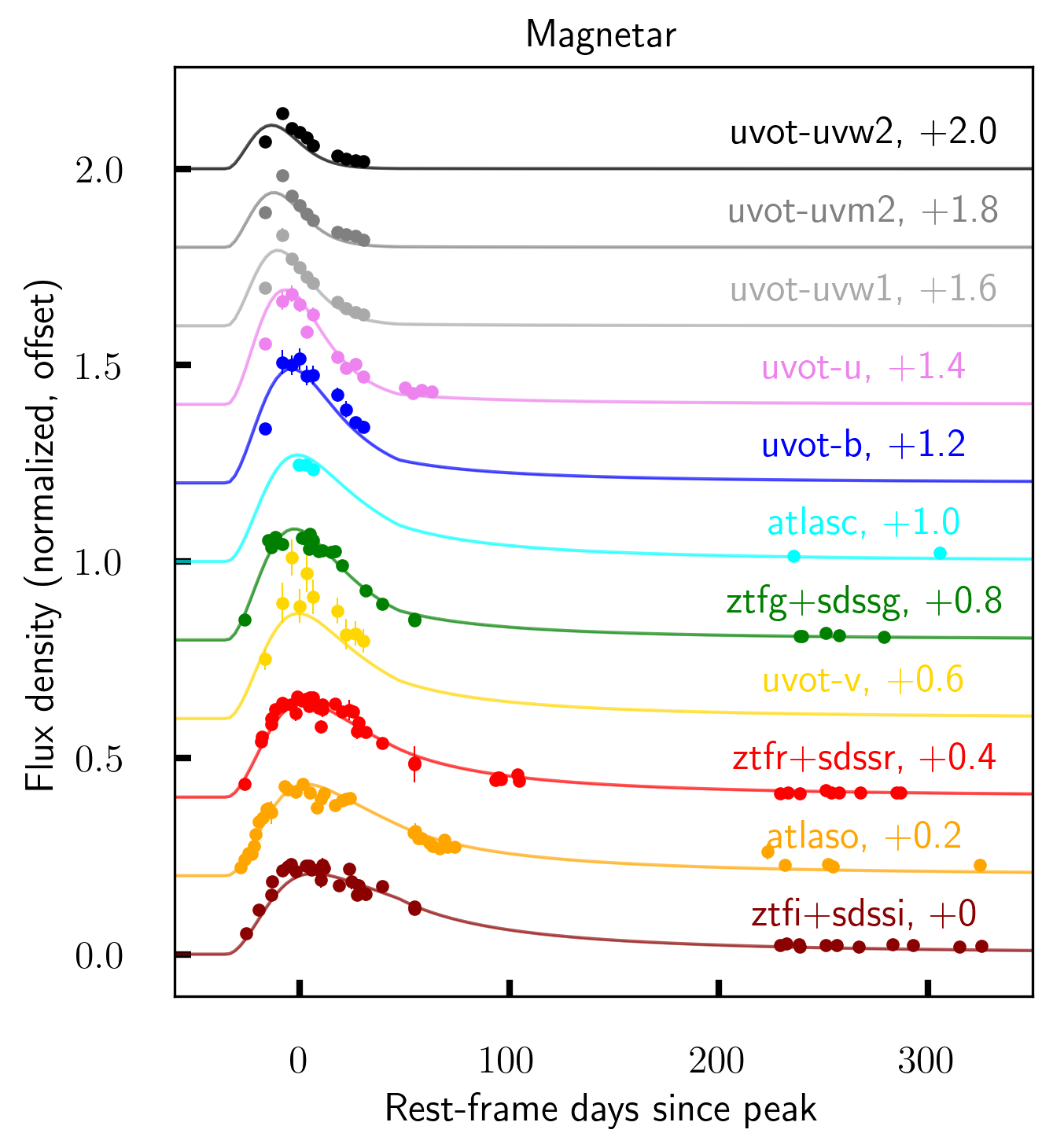}
    \caption{Fitting the multi-band photometry of SN\,2021aaev with the magnetar model from \citealt{Nicholl2017}. The fitting was implemented using \program{Redback} \citep{Sarin2024}.}
    \label{fig:redback}
\end{figure}

The spin-down of a magnetar is another commonly invoked powering mechanism to explain the light curve of SLSNe. We fitted the light curve of SN\,2021aaev with a magnetar model from \cite{Nicholl2017}. This model is available in \program{Redback} \citep{Sarin2024} as the \texttt{slsn} model, which is a magnetar model that fits multi-band light curve data with the constraints that the magnetar rotational energy is larger than the total output energy and that the nebula phase does not begin till at least 100 days. We performed nested sampling with default priors: $M_{\text{ej}}\in\log U[0.1,100]\,M_{\odot}$, $v_{\text{ej}}\in\log U[0.1,100]10^3$\,km\,s$^{-1}$, $P_0\in U[1,10]\,\text{ms}$ and $B_p\in\log U[0.1,10]\,10^{14}$\,G. The best-fit parameters are listed in \Cref{tab:models_parameter} and the comparison with observed multi-band data is shown in \Cref{fig:redback}.

The best-fit magnetar model reproduces the overall amplitude and trend of the $groi$ light curves but fails to capture the post-peak small-scale aperiodic fluctuations. The fit is notably poorer in the bluer bands, particularly at early phases. We infer a spin period $P_0$ of 3.95 ms and a magnetic field strength of $1.53\times10^{14}$ G, both of which are consistent with the reported values in the literature for SLSNe-I (e.g. $\sim2.5$ ms and $0.2-18\times10^{14}$ G, \citealt{Nicholl2017}). The inferred ejecta mass of $1.17\,M_{\odot}$ is however much lower than the median ejecta mass ($5.03^{+4.01}_{-2.39}$) of magnetar-powered SLSNe in \cite{Chen_2023_2}, but still within the plausible range. Hence, we conclude that the spin-down of a magnetar remains a viable mechanism for powering the generic light curve evolution of SN\,2021aaev. However, magnetar-powering alone cannot account for the aperiodic bumpy optical light curves or the persistent narrow emission lines seen in the spectra.

\section{Host galaxy environment} \label{sec:host_environment}

\subsection{Host SED modelling} \label{sec:host_SED}

%start from here
The SED and physical properties of the underlying host can provide further constraints on the progenitor and environment of SN2021aaev \citep[e.g.,][]{2021ApJS..255...29S}. An RGB composite of the host galaxy of SN\,2021aaev is shown in \Cref{fig:field}, and appears to be a spiral galaxy, with the supernova coincident with a red substructure in the south-eastern part (refer to as clump, hereafter). 

\begin{figure}[t]
    \centering
    \includegraphics[width=\linewidth]{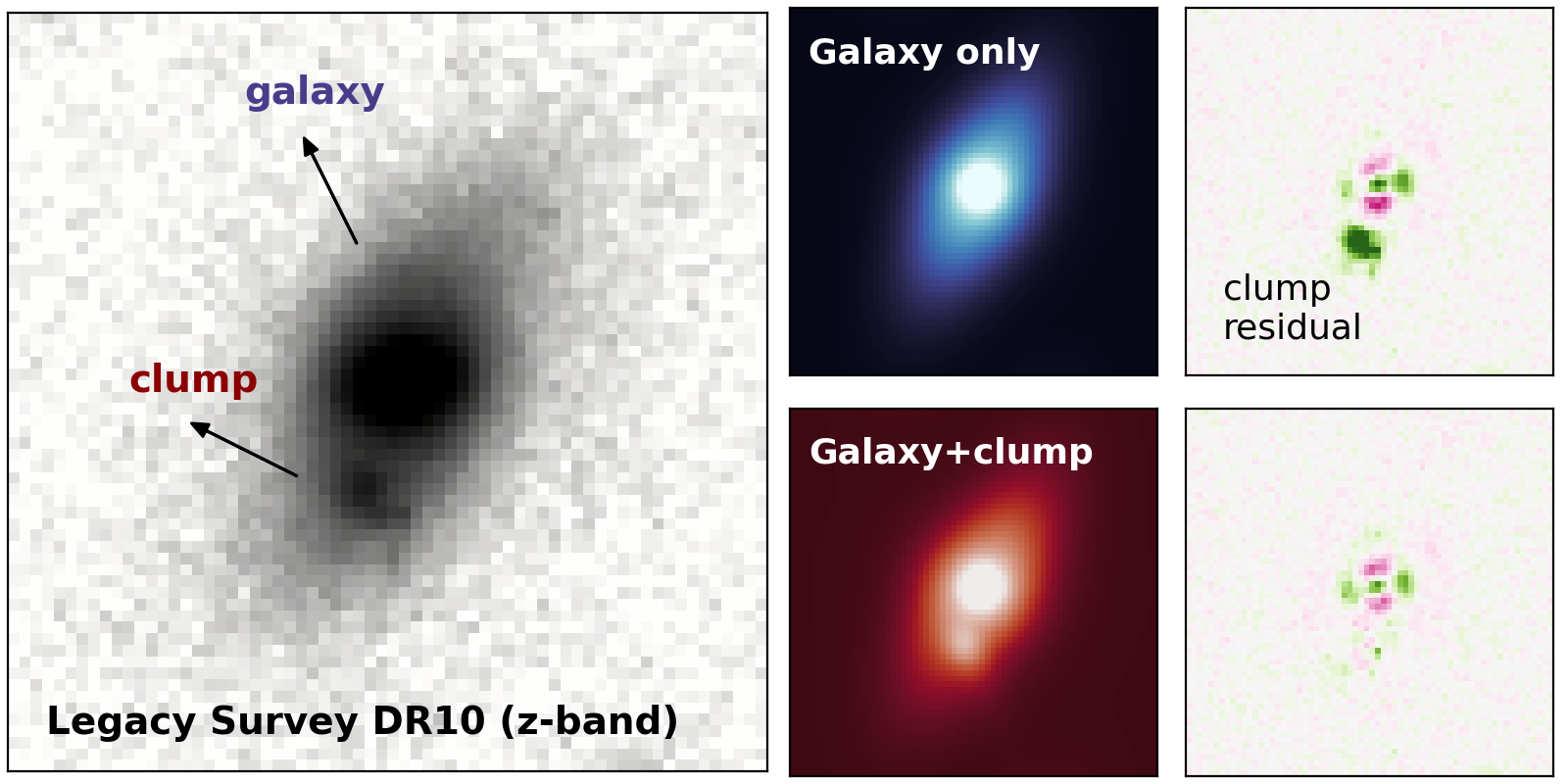}
    \caption{Morphological modeling of the SN\,2021aaev host. The \emph{left} panel shows the Legacy Survey DR10 z-band image of the system ($15\times15$~arcsec), highlighting the location of the central spiral galaxy and the accompanying clump, whose position coincides with the SN\,2021aaev explosion. The middle panels plot the best-fit \textsc{pysersic} models when only the galaxy (upper) or the galaxy+clump morphologies (lower) are considered. The \emph{right} panels show the residual of each fit: the presence of clump is clearly revealed in the upper plot.}
    \label{fig:host_clump_morph}
\end{figure}

We opt to decompose and study the SED of the spiral galaxy and the clump separately. First, we use the Bayesian Inference code \textsc{pysersic} \citep{Pasha2023} to characterize the morphology of the galaxy+clump system. We model the central galaxy using a double S{\'e}rsic profile with coincident centers and position angles, but varying effective radius, S{\'e}rsic index and ellipticity, aiming to better capture the 2D light distribution of the bulge and the spiral arms altogether. The southern clump is modeled simultaneously to the spiral using a point source. This procedure is applied to the \emph{griz} images of the system available from the Legacy Survey DR10. In order to better constrain the current SFR of the system, we also consider the \emph{u}-band exposures from SDSS/DR18 (although we acknowledge that the clump is not visible in the latter and, accordingly, a 5$\sigma$ limit on the flux is reported for this filter). After all, this method allow us to accurately extract both the photometry and morphology of the galaxy and the clump at the same time (see \Cref{fig:host_clump_morph}). The residual images reveal that the clump is resolved in the \emph{iz} photometry. In consequence, we model the morphology of  the clump in these two bands using a S{\'e}rsic profile. This results in a effective (or half-light) radius for the clump of $r_{\rm eff}^{\rm ~clump} = 3.8 \pm 0.3$ arcsec ($2.7 \pm 0.2$ kpc) in the \emph{z}-band, around three times smaller than the extent of the spiral companion of $r_{\rm eff}^{\rm ~galaxy} = 12.9 \pm 0.1$ arcsec ($9.0 \pm 0.1$ kpc). 

Then, we derived constraints on the SED properties of the galaxy and the clump (namely stellar masses, $M_{\star}$, and star-formation histories, SFHs) by fitting the available ground-based photometry with the Bayesian Analysis of Galaxies for Physical Inference and Parameter EStimation code, or \textsc{bagpipes} \citep{Carnall2018}. \textsc{bagpipes} uses the updated \citet{BC03} stellar population synthesis templates with a \citet{Kroupa2001} stellar initial mass function. In fitting the data, we adopt an exponentially declining prior for the SFH. We allowed for a broad range of stellar masses, $e$-folding times for the burst duration, and age interval limited to the age of the Universe at the redshift in question. For simplicity, and given the modest number of photometric points, the metallicity is fixed to the solar abundance value. We use a \citet{Calzetti2000} dust-attenuation law with the $V$-band optical depth allowed to vary between $0 \leq A_V \leq 2$ mag. 

\begin{figure}[t]
    \centering
    \includegraphics[width=\linewidth]{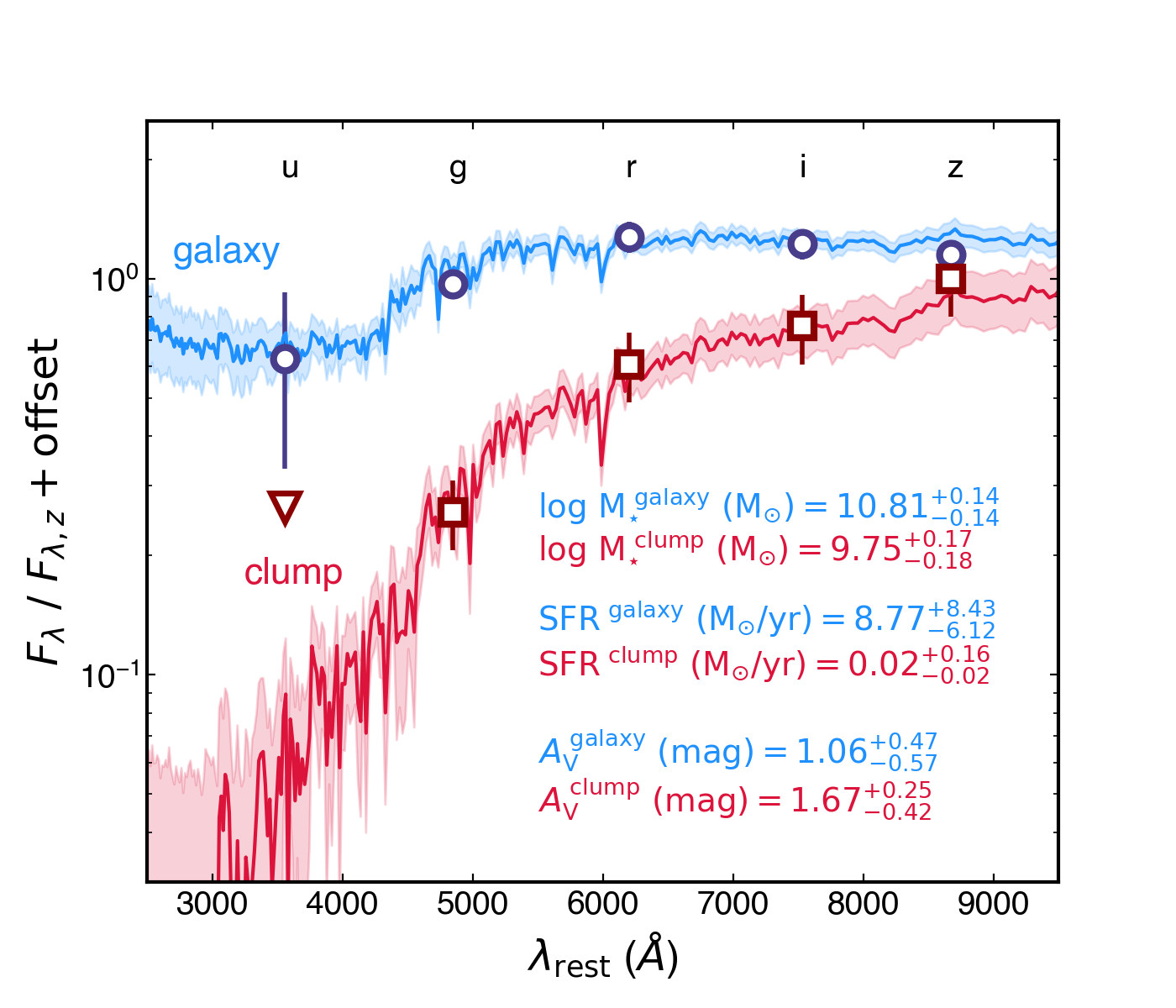}
    \caption{SED decomposition and physical properties of the system hosting SN\,2021aaev. The photometry and SED fits for both the spiral galaxy and the underlying clump are shown in blue and red, respectively, with the shaded band encompassing 16th to 84th percentile of the best-fit SEDs drawn from the \textsc{Bagpipes} realizations. The mean stellar mass, SFR and optical extinction are listed in the inset for each component (with uncertainties).}
    \label{fig:host_clump_sed}
\end{figure}

The observed photometry, best-fit SEDs and derived physical properties for the galaxy and clump are presented in \Cref{fig:host_clump_sed}. With a stellar mass of $\log(M_{\star}^{\rm clump}/{\rm M_{\odot}})\simeq 9.8$, the clump contributes more than 10\% to the total mass of the system, and it is significantly redder ($A_V^{\rm ~clump} \simeq 1.7 {\rm ~mag.}$), while the galaxy is compatible with lower extinction ($A_V^{\rm ~galaxy} \simeq 0.9 {\rm ~mag.}$) and a higher stellar mass of $M_{\star}^{\rm galaxy} \simeq 10.6~{\rm M_{\odot}}$. Remarkably, the spiral galaxy is still forming stars at an active pace (${\rm SFR} \simeq 5.47 {\rm ~M_{\odot}/yr}$), while the clump ceased forming stars long time ago. Given the clump's substantial mass, distinct SED and size compared to the main galaxy, the clump may represent a dwarf satellite galaxy or a merging companion with an old stellar population.

\subsection{Host galaxy lines}\label{sec:host}
\begin{figure*}[t]
    \centering
    \includegraphics[width=\linewidth]{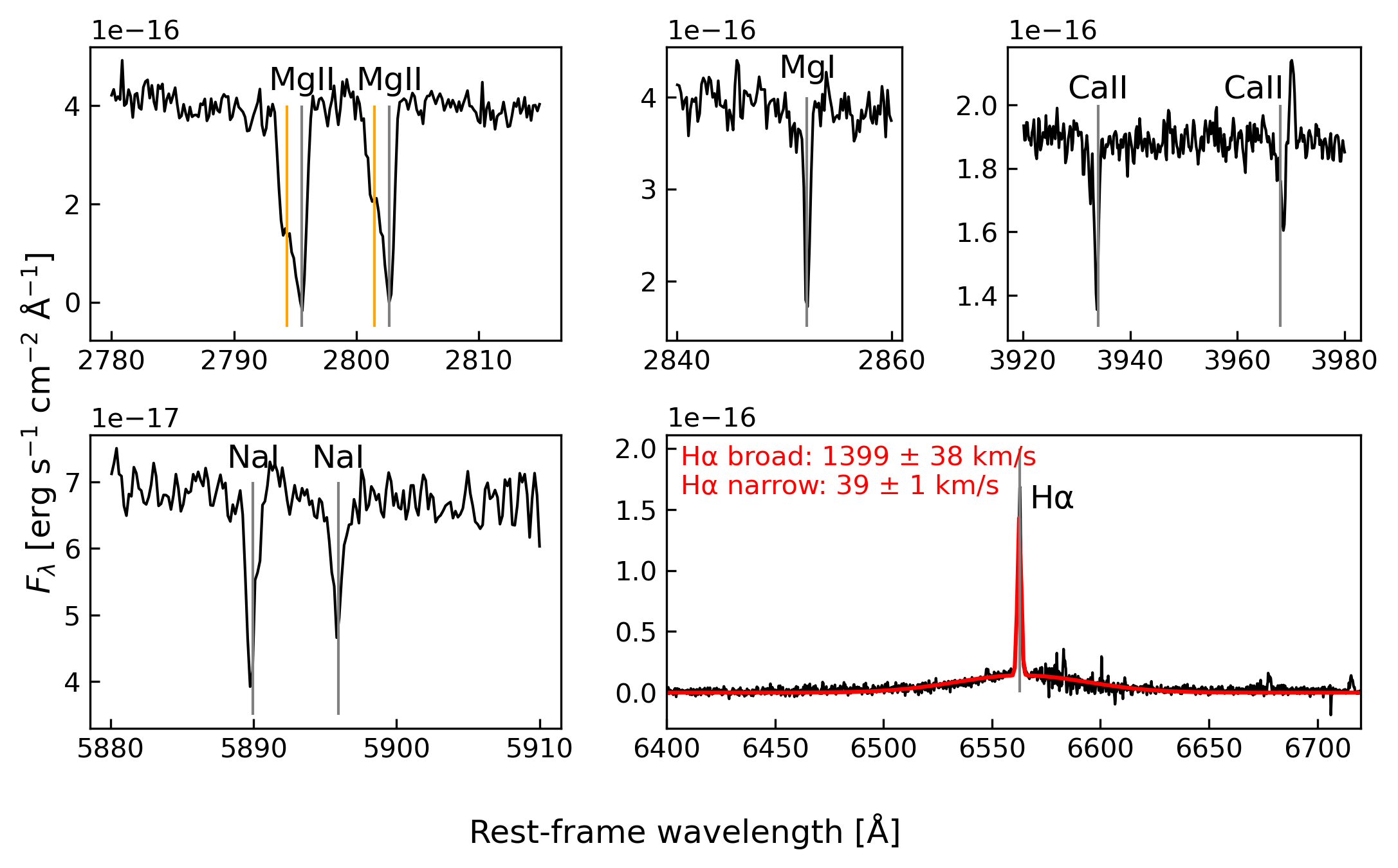}
    \caption{Potential host lines seen in the rest-frame $-16.7$ days X-Shooter spectrum of SN\,2021aaev. We identified host \ion{Mg}{1}, \ion{Mg}{2}, \ion{Ca}{2} and \ion{Na}{1} lines. The grey vertical lines are consistent with a redshift of $z=0.1557\pm0.0001$, and the orange lines have a relative velocity of $139\pm21$\,km\,s$^{-1}$ compared to the grey lines. A narrow component of H$\alpha$ at $z=0.1557\pm0.0001$ was seen, but further analysis points towards interaction with CSM rather than emissions from a host \ion{H}{2} region.}
    \label{fig:host_lines}
\end{figure*}

We observed the \ion{Mg}{2}~$\lambda\lambda$2797, 2803 doublet (saturated), the \ion{Mg}{1} $\lambda$2852 line, the \ion{Ca}{2} H ($\lambda$3968) and K ($\lambda$3934) lines, and the \ion{Na}{1} D $\lambda\lambda$5890, 5896 doublet in the $-16.7$ days X-Shooter spectrum, shown in \Cref{fig:host_lines}. These lines are commonly associated with absorption in the interstellar medium (ISM) of the host galaxy, and all of them are consistent with a redshift of $z=0.1557\pm0.0001$. Interestingly, the \ion{Mg}{2} doublet exhibit secondary blueshifted components, corresponding to a velocity offset of $139\pm21$\,km\,s$^{-1}$. This is consistent with the two-component host environment that we see in \Cref{fig:field} and the interpretation that the clump may be a dwarf satellite galaxy or a merging companion.

The equivalent width (EW) of the \ion{Na}{1} D $\lambda\lambda$5890, 5896 doublet is related to galaxy extinction. Using the empirical relation from \cite{NaID}, we derive a MW extinction of $E(B-V)_{\text{MW}} = 0.033 \pm 0.012$~mag and a host extinction of $E(B-V)_{\text{host}} = 0.086 \pm 0.032$~mag, leading to a total extinction of $E(B-V) = 0.119 \pm 0.044$~mag. The MW extinction is consistent with the value from NED's extinction calculator within the uncertainty (see \Cref{sec:correction}). While the inferred host reddening is non-negligible, the relation assumes that the extinction comes from a diffusive interstellar source rather than circumstellar dust. In the case of SN\,2021aaev, a SLSN-II with clear signs of CSM interaction, the local environment may also contain dusty CSM, which can contribute to the overall reddening in a way not captured by this relation. Given this ambiguity, we opt not to apply a host-extinction correction for photometric analysis. Nevertheless, the host extinction corresponds to $A_V=0.27$~mag, which is significantly lower than the value derived from the clump SED modelling in \Cref{sec:host_SED}. This suggests that the SN exploded in front of the clump along the line of sight. This raises a second possibility that the star-forming spiral in \Cref{fig:field} may have a foreground component that is in front of the red clump, and that the SN may have exploded in there. 

The narrow H lines observed in the X-Shooter spectra can resemble those originating from the host galaxy \ion{H}{2} regions. To investigate this, we analyzed the velocity dispersions of the \ion{Mg}{1}, \ion{Ca}{2}, and \ion{Na}{1} lines, finding values around 30\,km\,s$^{-1}$. The narrow components of H$\alpha$ gives a velocity dispersion of $39\pm1$\,km\,s$^{-1}$, and the H$\beta$ lines exhibit velocity dispersions of $29\pm2$\,km\,s$^{-1}$. Given that the resolution of the X-shooter VIS arm for the 0.9'' slit width is $R=8900$, these emission features are all resolved and are consistent with typical velocity dispersions observed in extragalactic \ion{H}{2} regions \citep{HII}. However, we did not observe strong [\ion{O}{3}] or [\ion{O}{2}] lines in proportion to the H lines. The Balmer decrement in the X-Shooter spectrum gives a value of 1.4 and is inconsistent with that from a typical \ion{H}{2} region. Considering these discrepancies with \ion{H}{2} region emissions, along with other evidence indicating the presence of CSM (see \Cref{sec:flash}), we conclude that the majority of the narrow component of Balmer lines in X-Shooter are attributed to CSM rather than the host.

\section{Discussion} \label{sec:dis}

\subsection{What CSM properties make SN\,2021aaev superluminous?}
SN\,2021aaev stands out as an extremely luminous H-rich interacting SN, with a peak luminosity three orders of magnitude brighter than some SNe IIn such as SN\,1998S, which is also thought to be primarily powered by interaction with CSM. This raises a key question: what CSM properties enable SN\,2021aaev to become superluminous?

The most important physical parameter that characterize the ejecta-CSM interaction is perhaps the mass ratio between CSM and ejecta. From SN\,2021aaev, we saw light-curve and spectral evidence of long-lived CSM interaction, with no identifiable ejecta features for up to about 100 days since first light. This implies that the ejecta were constantly buried within an optically thick CSM and have experienced significant deceleration. The degree of deceleration is closely related to the mass ratio between CSM and ejecta. This parameter also governs the energy conversion efficiency, where a larger $M_{\text{CSM}}:M_{\text{ej}}$ generally gives a higher efficiency. However, energy conversion efficiency does not necessarily increase monotonically with CSM mass for a given ejecta kinetic energy. As argued by \citet{khatami2024}, there may exist an optimal mass ratio near unity that maximizes radiative output. Beyond this point, increasing the CSM mass merely decelerate the ejecta. In the case of SN\,2021aaev, which has an inferred CSM-ejecta mass ratio of $\sim$9, the conversion efficiency may already be somewhat suppressed relative to this optimum. Nonetheless, our analytical model in \Cref{sec:model} demonstrates that a CSM that outmasses the ejecta provides a viable explanation for the pseudo-bolometric light curve of SN\,2021aaev. This scenario also naturally extends to other luminous, slowly evolving H-rich interacting SNe with comparable peak luminosities and timescales, such as SN\,2017hcc, SN\,2006gy and SN\,2010jl. In contrast, events such as SN\,1998S may involve lower $M_{\text{CSM}}:M_{\text{ej}}$, resulting in less efficient energy conversion and faster light-curve evolution. Still, a handful of well-studied cases is insufficient to answer whether SLSNe-IIn are merely the luminous tail of the SN IIn population. A more systematic, sample-based approach, incorporating both photometry and spectroscopy, is needed to address this question robustly.

Another key parameter is the radial extent of the CSM. In SN\,2021aaev, spectroscopic evidence points to a complex, extended and perhaps stratified CSM configuration. The coexistence of persistent narrow Balmer emission lines and early-time flash-ionization features with broad Lorentzian wings suggests a radially layered CSM. Specifically, the transient flash features around 4650 \AA{}, attributed to He, N, and/or C, imply the presence of a compact, dense inner shell. This is a phase where interaction physics may deviate from other phases (see e.g. \citealt{matsuoka2025}), while the persistent narrow Balmer components arise from more extended, photo-ionized H-rich material that remains unshocked. This structure likely reflects a history of episodic pre-SN mass-loss events, transitioning from a continuous, wind-driven outflow to more explosive or eruptive ejections closer to core collapse.

Given this inferred massive, extensive CSM, what can we say about the pre-explosion activities and the progenitor of SN\,2021aaev? The inferred CSM mass is comparable to that of SN\,2006gy, which required a $\sim 10$--$20\,M_{\odot}$ CSM to explain its luminosity and duration \citep[e.g.][]{Smith2010,2006gy2}. We can estimate the progenitor mass-loss rate by assuming $\dot{M}=4\pi v_{\text{w}}\rho_{\text{CSM}}R^2_{\text{CSM}}$ with wind velocity $v_{\mathrm{w}}=10$--$100$\,km\,s$^{-1}$. This gives a mass-loss rate of $0.1$--$0.8\,M_{\odot}\,\text{yr}^{-1}$ for CSM density parameter $s=2$, which is comparable to that of the SN\,2006gy ($0.1\,M_{\odot}\,\text{yr}^{-1}$, \citealt{2006gy_0}) and the extreme value inferred for SN IIn events like iPTF13z, which showed $\dot{M} \sim 0.1$--$2\,M_{\odot}\,\text{yr}^{-1}$ \citep{Nyholm2017}. This estimated mass-loss rate is consistent with extreme eruptive episodes such as those seen in luminous blue variables (LBVs) undergoing giant eruptions \citep[see e.g.][]{2006ApJ...645L..45S}, or the violent mass ejections predicted by pulsational pair-instability events in very massive stars \citep{2017ApJ...836..244W}. This supports the idea that SN\,2021aaev's progenitor may have undergone similarly eruptive pre-supernova mass loss.

\subsection{Can SN\,2021aaev be a SN Ia-CSM?}

While CSM interaction dominates the observed emission of SN\,2021aaev, it also obscures the nature of the underlying explosion. Moreover, the transient location's visual alignment with a low-SFR region ($0.02~M_\odot$\,yr$^{-1}$) raises the possibility of an origin other than massive star core-collapse. Here we discuss whether SN\,2021aaev could be a Type Ia supernova interacting with CSM (SN Ia-CSM), i.e. a thermonuclear explosion embedded in a dense CSM envelope.

SNe Ia-CSM can sometimes be disguised as SNe IIn or SLSNe-IIn \citep{Ia-CSM3}, since SN Ia-CSM can reach comparable peak luminosities (as bright as $-21.3$ mag in the $R$ band, \citealt{Ia-CSM1}) and show prominent H$\alpha$. Systematic sample studies \citep{Ia-CSM1, Ia-CSM2} have shown that the spectra of the less luminous SNe Ia-CSM typically exhibit ``diluted'' Ia features blended with H emissions and a Balmer decrement of $5$-$7$. Although such features are not observed in the early-time spectra of SN\,2021aaev, it is worth noting, as \cite{Ia-CSM3} pointed out, that in more luminous cases ($<-20.5$\,mag), the underlying SN Ia features can remain hidden until late phases, especially in cases where $M_{\rm CSM} \gg M_{\rm ej}$. In the absence of late-time spectra, we cannot completely rule out the SN Ia-CSM scenario based on current observations.

Notably, \citet{2006gy_3} suggested that SN\,2006gy, which was long considered the ``prototypical'' SLSN-IIn, could be explained by a SN Ia-CSM model based on its high inferred nickel mass ($0.5\,M_\odot$) and intermediate-width Fe lines in the $+394$ day spectrum. Despite the spectroscopic differences, SN\,2021aaev shares several similarities with SN\,2006gy. Firstly, they exhibit comparable peak luminosities, total radiated energies, and light curve timescales. Secondly, from our simple analytical modelling in \Cref{sec:model}, we inferred an ejecta mass of $1$-$2\,M_\odot$ that does not rule out a thermonuclear origin, and a CSM mass of $\sim13\,M_\odot$ that matches the CSM estimates for SN\,2006gy in \cite{2006gy_3}. As \cite{2006gy_3} proposed, one plausible progenitor scenario capable of producing such a large CSM mass is the inspiral of a white dwarf (WD) into a massive companion with a non-degenerate helium core, which can potentially unbind a substantial fraction of the stellar envelope during the process \citep[simulated with a neutron star; similar dynamics are expected for a WD;][]{Ia-CSM4}.

\subsection{Can SN\,2021aaev be a nuclear transient?}
If we interpret the red clump in \Cref{fig:field} that SN\,2021aaev visually aligns with as a distinct galaxy, likely a dwarf galaxy interacting with a larger spiral, then we should also consider the possibility of SN\,2021aaev being of nuclear origin. One class of nuclear transients is tidal disruption events (TDEs), which can be contaminants in SLSN-II samples \citep[see e.g.][]{pessi2025} due to their similarly extreme luminosities and energetics. This is especially relevant in light of the recent study on AT\,2022wtn \citep{AT2022wtn}, where a TDE was discovered in an interacting galaxy environment, and AT\,2020yue, which was initially considered to be a SLSN-II \citep{Kangas} but later reclassified as a TDE by \cite{Yao_2023_TDE}. Yet, despite the similar location, SN\,2021aaev shows distinct photometric and spectroscopic properties compared to TDEs. TDEs typically have little color evolution and maintain a constant or even rising blackbody temperature over time \citep{TDE1,TDE2}, while SN\,2021aaev evolves from blue to red and cools after peak, which is more consistent with a SN behavior. The spectra of SN\,2021aaev evolve on a timescale of tens of days, while TDEs typically show much slower spectral evolution, often remaining blue and only weakly evolving for several months \citep{TDE2,TDE3}. Hence, we conclude that the observations of SN\,2021aaev agrees better with an interacting SN rather than a TDE.

Can SN\,2021aaev be a changing-look active galactic nucleus (CLAGN)? CLAGN is a subclass of AGNs that change types due to either changes in accretion rate or obscuration \citep[see e.g.][]{CLAGN1}. Owing to their diverse observable properties, some CLAGNs can mimic SLSNe-IIn. A notable example is AT\,2022rze \citep{AT2022rze}, which is an ambiguously classified transient found in a merging system. However, SN\,2021aaev shows smooth, well-behaved photometric and spectroscopic evolution, while CLAGN variability is generally stochastic and persistent. Many CLAGNs display a ``turn-on'' of broad, persistent Balmer emissions that may arise from accretion-state changes \citep{CLAGN2}, and show typical AGN lines such as [\ion{O}{3}] $\lambda\lambda$4959,5007, [\ion{N}{2}] $\lambda\lambda$6548,6583 and [\ion{S}{2}] $\lambda\lambda$6716,6732 \citep{CLAGN3}. None of these is observed in the case of SN\,2021aaev. Therefore, SN\,2021aaev is better interpreted as an interacting SN rather than a CLAGN.

\section{Conclusion} \label{sec:con}
In this work, we have presented and analyzed the photometric and spectroscopic dataset of SN\,2021aaev, a hydrogen-rich, superluminous, interacting SN discovered by ZTF and classified as a SLSN-IIn. Our main conclusions are as follows:
\begin{enumerate}[itemsep=0pt, parsep=0pt, topsep=0pt]
    \item SN\,2021aaev lies at the luminous ($o$-band peak at $-21.35$~mag), fast-evolving ($t_{\text{rise}}=32.5$~days) side of the SLSN-II population, and has a total radiated energy of $1.41\times10^{51}$~erg.
    \item SN\,2021aaev shows persistent narrow Balmer lines for $\sim100$ days, likely explained by continued CSM interaction.
    \item Early spectra reveal a fading feature at $\sim4650$ \AA{} that we identify as a blend of flash-ionized \ion{He}{2}, \ion{C}{3} and/or \ion{N}{3} lines from dense, confined CSM, making this the first known SLSN-IIn flasher.
    \item Analytical modelling favors a scenario where SN\,2021aaev is powered by massive, extensive H-rich CSM ($M_{\text{CSM}}\sim9$--$19\,M_{\odot}$, $R_{\text{CSM}}\sim1.3$--$2.0\times10^{16}$~cm) that outmasses the ejecta ($M_{\text{ej}}\sim1$--$2\,M_{\odot}$), likely from eruptive LBV or pulsational pair-instability episodes. Alternatively, as the massive CSM covers up the true nature of the explosion, a Type Ia-CSM origin cannot be ruled out.
    \item SN\,2021aaev is visually located on a red substructure (likely a dwarf satellite or a merging companion) within a larger spiral host. The substructure's SED and absence of strong emission lines indicate a quiescent environment (SFR$=0.02^{+0.13}_{-0.02}\,M_{\odot}$~yr$^{-1}$).
\end{enumerate}

Overall, SN\,2021aaev highlights the challenges in disentangling the underlying explosion mechanisms of SLSNe-IIn from strong CSM interaction. Its combination of early-time flash features, massive CSM, and unusual host environment adds to the observed diversity among H-rich SLSNe.

\begin{acknowledgments}
\footnotesize
\textit{Acknowledgments.} 
We thank Daniel A. Perley for the observation made with the IO:O instrument on the Liverpool Telescope (LT), and K-Ryan Hinds for the reduction on the photometry taken with the LT. We thank Luc Dessart, Claes Fransson and Nikhil Sarin for useful discussions. 
T.-W.C. acknowledges the financial support from the Yushan Fellow Program by the Ministry of Education, Taiwan (MOE-111-YSFMS-0008-001-P1) and the National Science and Technology Council, Taiwan (NSTC grant 114-2112-M-008-021-MY3). MN is supported by the European Research Council (ERC) under the European Union's Horizon 2020 research and innovation programme (grant agreement No.\~948381).
The Oskar Klein Centre is funded by the Swedish Research Council. 
This project is funded by the European Union (ERC, project number 101042299, TransPIre). Views and opinions expressed are however those of the author(s) only and do not necessarily reflect those of the European Union or the European Research Council Executive Agency. Neither the European Union nor the granting authority can be held responsible for them. 
Based on observations obtained with the Samuel Oschin Telescope 48-inch and the 60-inch Telescope at the Palomar Observatory as part of the Zwicky Transient Facility project. ZTF is supported by the National Science Foundation under Grants No. AST-1440341, AST-2034437, and currently Award \#2407588. ZTF receives additional funding from the ZTF partnership. Current members include Caltech, USA; Caltech/IPAC, USA; University of Maryland, USA; University of California, Berkeley, USA; University of Wisconsin at Milwaukee, USA; Cornell University, USA; Drexel University, USA; University of North Carolina at Chapel Hill, USA; Institute of Science and Technology, Austria; National Central University, Taiwan, and OKC, University of Stockholm, Sweden. Operations are conducted by Caltech's Optical Observatory (COO), Caltech/IPAC, and the University of Washington at Seattle, USA. SED Machine is based upon work supported by the National Science Foundation under Grant No. 1106171. 
Based on observations collected at the European Southern Observatory under ESO program IDs 106.216C and 108.220C (PI: Inserra). 
This work has made use of  data from ALFOSC, which is provided by the instituto de Astrofisica de Andalucia (IAA) under a joint agreement with the University of Copenhagen and NOT. The ZTF forced-photometry service was funded under the Heising-Simons Foundation grant \#12540303 (PI: Graham).
This work has made use of data from the Asteroid Terrestrial-impact Last Alert System (ATLAS) project. ATLAS is primarily funded to search for near earth asteroids through NASA grants NN12AR55G, 80NSSC18K0284, and 80NSSC18K1575; byproducts of the NEO search include images and catalogs from the survey area. The ATLAS science products have been made possible through the contributions of the University of Hawaii Institute for Astronomy, the Queen's University Belfast, the Space Telescope Science Institute, and the South African Astronomical Observatory. 
The Liverpool Telescope is operated on the island of La Palma by Liverpool John Moores University in the Spanish Observatorio del Roque de los Muchachos of the Instituto de Astrofisica de Canarias with financial support from the UK Science and Technology Facilities Council.
\end{acknowledgments}

\vspace{5mm}
\facilities{ZTF, SEDM, KAIT, LT, ATLAS, \textit{Swift}(UVOT), ESO VLT, NOT, ESO NTT}
\software{\textsc{Numpy} \citep{van_der_Walt_2011_Numpy}, \textsc{SciPy} \citep{Virtanen_2020_SciPy}, \textsc{Matplotlib} \citep{Hunter_2007_Matplotlib}, \textsc{astropy} \citep{2013A&A...558A..33A,2018AJ....156..123A}, \textsc{george} \citep{2015ascl.soft11015F}, \textsc{redback} \citep{Sarin2024}, \textsc{emcee} \citep{2013PASP..125..306F}, \textsc{bagpipes} \citep{Carnall2018}, \textsc{pysersic} \citep{Pasha2023}}, and the Fritz SkyPortal Marshal \citep{Fritz_1, Fritz_2}

\clearpage

\appendix
\section{Photometric Data for SN~2021aaev}\label{appendix:photometry}
\twocolumngrid
\input{appendix/SN2021aaev_photometry_table.txt}

\clearpage
\section{Spectroscopic Data for SN~2021aaev}\label{appendix:spectroscopy}
\input{appendix/SN2021aaev_spectroscopy_table.txt}

\clearpage
\onecolumngrid
\section{Result from Analytical CSM-interaction Models}\label{appendix:MCMC}
\input{appendix/MCMC.txt}
\clearpage
\bibliography{ref}{}
\bibliographystyle{aasjournal}

\end{document}

%% file: appendix/SN2021aaev_photometry_table.txt
\startlongtable
\begin{deluxetable}{lcccc}
\tablecaption{Photometric observations of SN~2021aaev. \label{tab:photometry}}
\tablehead{
\colhead{MJD} & \colhead{Phase\tablenotemark{a}} & \colhead{Filter} & \colhead{Magnitude} & \colhead{Instrument} \\
\colhead{} & \colhead{[day]} & \colhead{} & \colhead{[mag]} & \colhead{}
}
\startdata
59486.43 & -31.57 & $o$ & $20.595 \pm 0.269$ & ATLAS \\
59488.33 & -29.67 & $r$ & $20.105 \pm 0.112$ & ZTF \\
59488.38 & -29.62 & $g$ & $19.629 \pm 0.070$ & ZTF \\
59488.51 & -29.49 & $o$ & $19.868 \pm 0.182$ & ATLAS \\
59489.30 & -28.70 & $i$ & $19.592 \pm 0.146$ & ZTF \\
59490.51 & -27.49 & $o$ & $19.520 \pm 0.116$ & ATLAS \\
59492.42 & -25.58 & $o$ & $19.559 \pm 0.211$ & ATLAS \\
59493.44 & -24.56 & $o$ & $19.216 \pm 0.086$ & ATLAS \\
59494.44 & -23.56 & $o$ & $18.842 \pm 0.066$ & ATLAS \\
59496.34 & -21.66 & $i$ & $18.772 \pm 0.067$ & ZTF \\
59496.42 & -21.58 & $o$ & $18.558 \pm 0.062$ & ATLAS \\
59497.43 & -20.57 & $r$ & $18.520 \pm 0.043$ & ZTF \\
59498.15 & -19.85 & $r$ & $18.441 \pm 0.071$ & SEDM \\
59498.47 & -19.53 & $o$ & $18.480 \pm 0.046$ & ATLAS \\
59499.67 & -18.33 & $V$ & $18.452 \pm 0.193$ & $Swift$/UVOT \\
59499.67 & -18.33 & $UVW2$ & $19.313 \pm 0.087$ & $Swift$/UVOT \\
59499.67 & -18.33 & $UVW1$ & $18.940 \pm 0.089$ & $Swift$/UVOT \\
59499.67 & -18.33 & $B$ & $18.555 \pm 0.114$ & $Swift$/UVOT \\
59499.67 & -18.33 & $U$ & $18.432 \pm 0.084$ & $Swift$/UVOT \\
59499.68 & -18.32 & $UVM2$ & $19.026 \pm 0.091$ & $Swift$/UVOT \\
59500.38 & -17.62 & $o$ & $18.322 \pm 0.059$ & ATLAS \\
59501.40 & -16.60 & $g$ & $17.890 \pm 0.018$ & ZTF \\
59501.40 & -16.60 & $o$ & $18.321 \pm 0.048$ & ATLAS \\
59503.21 & -14.79 & $i$ & $18.449 \pm 0.064$ & ZTF \\
59503.30 & -14.70 & $r$ & $18.151 \pm 0.039$ & ZTF \\
59503.32 & -14.68 & $o$ & $18.383 \pm 0.180$ & ATLAS \\
59503.34 & -14.66 & $g$ & $17.966 \pm 0.036$ & ZTF \\
59503.35 & -14.65 & $r$ & $18.225 \pm 0.046$ & SEDM \\
59503.39 & -14.61 & $r$ & $18.139 \pm 0.027$ & SEDM \\
59503.39 & -14.61 & $i$ & $18.230 \pm 0.036$ & SEDM \\
59503.39 & -14.61 & $g$ & $17.925 \pm 0.047$ & SEDM \\
59505.21 & -12.79 & $g$ & $17.854 \pm 0.048$ & ZTF \\
59505.29 & -12.71 & $r$ & $18.028 \pm 0.042$ & ZTF \\
59509.22 & -8.78 & $r$ & $17.949 \pm 0.033$ & SEDM \\
59509.25 & -8.75 & $g$ & $17.931 \pm 0.068$ & SEDM \\
59509.25 & -8.75 & $r$ & $17.988 \pm 0.053$ & SEDM \\
59509.26 & -8.74 & $i$ & $18.075 \pm 0.023$ & SEDM \\
59509.32 & -8.68 & $UVW1$ & $17.988 \pm 0.079$ & $Swift$/UVOT \\
59509.33 & -8.67 & $U$ & $17.847 \pm 0.096$ & $Swift$/UVOT \\
59509.33 & -8.67 & $UVM2$ & $18.248 \pm 0.073$ & $Swift$/UVOT \\
59509.33 & -8.67 & $V$ & $17.727 \pm 0.190$ & $Swift$/UVOT \\
59509.33 & -8.67 & $UVW2$ & $18.526 \pm 0.087$ & $Swift$/UVOT \\
59509.33 & -8.67 & $B$ & $17.682 \pm 0.108$ & $Swift$/UVOT \\
59510.42 & -7.58 & $o$ & $18.004 \pm 0.047$ & ATLAS \\
59512.25 & -5.75 & $i$ & $18.030 \pm 0.032$ & ZTF \\
59512.43 & -5.57 & $o$ & $18.044 \pm 0.047$ & ATLAS \\
59514.19 & -3.81 & $i$ & $18.001 \pm 0.058$ & SEDM \\
59514.19 & -3.81 & $r$ & $17.973 \pm 0.032$ & SEDM \\
59514.34 & -3.66 & $B$ & $17.709 \pm 0.088$ & $Swift$/UVOT \\
59514.34 & -3.66 & $UVW2$ & $18.869 \pm 0.083$ & $Swift$/UVOT \\
59514.34 & -3.66 & $U$ & $17.776 \pm 0.078$ & $Swift$/UVOT \\
59514.34 & -3.66 & $UVW1$ & $18.314 \pm 0.074$ & $Swift$/UVOT \\
59514.34 & -3.66 & $V$ & $17.368 \pm 0.123$ & $Swift$/UVOT \\
59514.35 & -3.65 & $UVM2$ & $18.603 \pm 0.069$ & $Swift$/UVOT \\
59516.41 & -1.59 & $r$ & $18.076 \pm 0.100$ & KAIT \\
59516.41 & -1.59 & $i$ & $18.090 \pm 0.104$ & KAIT \\
59516.41 & -1.59 & $o$ & $18.074 \pm 0.035$ & ATLAS \\
59517.27 & -0.73 & $r$ & $17.884 \pm 0.017$ & ZTF \\
59518.40 & 0.40 & $c$ & $17.921 \pm 0.025$ & ATLAS \\
59518.75 & 0.75 & $UVW1$ & $18.462 \pm 0.081$ & $Swift$/UVOT \\
59518.76 & 0.76 & $UVW2$ & $18.972 \pm 0.088$ & $Swift$/UVOT \\
59518.76 & 0.76 & $U$ & $17.888 \pm 0.085$ & $Swift$/UVOT \\
59518.76 & 0.76 & $V$ & $17.759 \pm 0.163$ & $Swift$/UVOT \\
59518.76 & 0.76 & $UVM2$ & $18.831 \pm 0.077$ & $Swift$/UVOT \\
59518.76 & 0.76 & $B$ & $17.651 \pm 0.091$ & $Swift$/UVOT \\
59520.27 & 2.27 & $r$ & $17.927 \pm 0.019$ & ZTF \\
59520.29 & 2.29 & $g$ & $17.858 \pm 0.017$ & ZTF \\
59520.41 & 2.41 & $o$ & $17.978 \pm 0.031$ & ATLAS \\
59522.27 & 4.27 & $g$ & $17.874 \pm 0.017$ & ZTF \\
59522.28 & 4.28 & $i$ & $18.014 \pm 0.025$ & ZTF \\
59522.30 & 4.30 & $r$ & $17.910 \pm 0.017$ & ZTF \\
59522.40 & 4.40 & $c$ & $17.919 \pm 0.027$ & ATLAS \\
59522.67 & 4.67 & $B$ & $17.809 \pm 0.100$ & $Swift$/UVOT \\
59522.67 & 4.67 & $UVW1$ & $18.662 \pm 0.087$ & $Swift$/UVOT \\
59522.67 & 4.67 & $U$ & $18.233 \pm 0.098$ & $Swift$/UVOT \\
59522.67 & 4.67 & $UVW2$ & $19.161 \pm 0.093$ & $Swift$/UVOT \\
59522.68 & 4.68 & $UVM2$ & $19.066 \pm 0.082$ & $Swift$/UVOT \\
59522.68 & 4.68 & $V$ & $17.481 \pm 0.140$ & $Swift$/UVOT \\
59523.39 & 5.39 & $r$ & $17.953 \pm 0.085$ & KAIT \\
59524.15 & 6.15 & $r$ & $17.984 \pm 0.021$ & SEDM \\
59524.18 & 6.18 & $r$ & $17.917 \pm 0.017$ & SEDM \\
59524.18 & 6.18 & $i$ & $18.015 \pm 0.022$ & SEDM \\
59524.18 & 6.18 & $g$ & $17.989 \pm 0.034$ & SEDM \\
59524.25 & 6.25 & $r$ & $17.886 \pm 0.018$ & ZTF \\
59524.32 & 6.32 & $g$ & $17.824 \pm 0.015$ & ZTF \\
59524.34 & 6.34 & $o$ & $18.087 \pm 0.040$ & ATLAS \\
59525.18 & 7.18 & $i$ & $18.069 \pm 0.036$ & ZTF \\
59526.25 & 8.25 & $UVW2$ & $19.481 \pm 0.100$ & $Swift$/UVOT \\
59526.25 & 8.25 & $r$ & $17.892 \pm 0.017$ & ZTF \\
59526.25 & 8.25 & $UVW1$ & $18.817 \pm 0.090$ & $Swift$/UVOT \\
59526.25 & 8.25 & $U$ & $18.005 \pm 0.087$ & $Swift$/UVOT \\
59526.25 & 8.25 & $V$ & $17.672 \pm 0.158$ & $Swift$/UVOT \\
59526.25 & 8.25 & $B$ & $17.802 \pm 0.096$ & $Swift$/UVOT \\
59526.32 & 8.32 & $UVM2$ & $19.293 \pm 0.135$ & $Swift$/UVOT \\
59526.32 & 8.32 & $g$ & $17.885 \pm 0.017$ & ZTF \\
59526.38 & 8.38 & $c$ & $17.974 \pm 0.024$ & ATLAS \\
59528.40 & 10.40 & $o$ & $18.297 \pm 0.063$ & ATLAS \\
59529.27 & 11.27 & $i$ & $18.063 \pm 0.037$ & ZTF \\
59529.27 & 11.27 & $g$ & $18.013 \pm 0.022$ & ZTF \\
59529.34 & 11.34 & $r$ & $18.009 \pm 0.026$ & ZTF \\
59530.36 & 12.36 & $r$ & $18.268 \pm 0.105$ & KAIT \\
59530.36 & 12.36 & $o$ & $18.178 \pm 0.121$ & ATLAS \\
59530.36 & 12.36 & $i$ & $18.207 \pm 0.114$ & KAIT \\
59531.24 & 13.24 & $r$ & $17.972 \pm 0.028$ & ZTF \\
59531.33 & 13.33 & $g$ & $18.006 \pm 0.029$ & ZTF \\
59531.36 & 13.36 & $r$ & $18.030 \pm 0.093$ & KAIT \\
59531.36 & 13.36 & $i$ & $18.009 \pm 0.089$ & KAIT \\
59532.15 & 14.15 & $i$ & $18.046 \pm 0.041$ & ZTF \\
59532.28 & 14.28 & $o$ & $18.097 \pm 0.071$ & ATLAS \\
59536.31 & 18.31 & $g$ & $18.027 \pm 0.059$ & ZTF \\
59538.20 & 20.20 & $g$ & $18.015 \pm 0.042$ & ZTF \\
59538.25 & 20.25 & $r$ & $17.960 \pm 0.029$ & ZTF \\
59538.44 & 20.44 & $o$ & $18.265 \pm 0.080$ & ATLAS \\
59539.72 & 21.72 & $U$ & $18.706 \pm 0.096$ & $Swift$/UVOT \\
59539.72 & 21.72 & $B$ & $18.026 \pm 0.088$ & $Swift$/UVOT \\
59539.72 & 21.72 & $UVW1$ & $19.453 \pm 0.095$ & $Swift$/UVOT \\
59539.72 & 21.72 & $UVW2$ & $20.083 \pm 0.103$ & $Swift$/UVOT \\
59539.72 & 21.72 & $V$ & $17.803 \pm 0.131$ & $Swift$/UVOT \\
59539.73 & 21.73 & $UVM2$ & $19.901 \pm 0.090$ & $Swift$/UVOT \\
59540.28 & 22.28 & $i$ & $18.294 \pm 0.073$ & ZTF \\
59542.18 & 24.18 & $r$ & $18.053 \pm 0.024$ & ZTF \\
59542.28 & 24.28 & $g$ & $18.209 \pm 0.043$ & ZTF \\
59542.33 & 24.33 & $o$ & $18.197 \pm 0.044$ & ATLAS \\
59544.37 & 26.37 & $o$ & $18.175 \pm 0.045$ & ATLAS \\
59544.50 & 26.50 & $B$ & $18.220 \pm 0.121$ & $Swift$/UVOT \\
59544.50 & 26.50 & $UVW1$ & $19.769 \pm 0.134$ & $Swift$/UVOT \\
59544.50 & 26.50 & $U$ & $18.993 \pm 0.137$ & $Swift$/UVOT \\
59544.51 & 26.51 & $V$ & $18.071 \pm 0.195$ & $Swift$/UVOT \\
59544.51 & 26.51 & $UVM2$ & $20.131 \pm 0.119$ & $Swift$/UVOT \\
59544.51 & 26.51 & $UVW2$ & $20.367 \pm 0.138$ & $Swift$/UVOT \\
59546.31 & 28.31 & $i$ & $18.054 \pm 0.077$ & KAIT \\
59546.31 & 28.31 & $r$ & $18.034 \pm 0.128$ & KAIT \\
59546.37 & 28.37 & $o$ & $18.164 \pm 0.037$ & ATLAS \\
59547.33 & 29.33 & $i$ & $18.225 \pm 0.121$ & KAIT \\
59548.32 & 30.32 & $r$ & $18.060 \pm 0.070$ & KAIT \\
59549.51 & 31.51 & $UVW1$ & $20.082 \pm 0.124$ & $Swift$/UVOT \\
59549.51 & 31.51 & $B$ & $18.432 \pm 0.108$ & $Swift$/UVOT \\
59549.51 & 31.51 & $UVW2$ & $20.550 \pm 0.121$ & $Swift$/UVOT \\
59549.51 & 31.51 & $U$ & $18.879 \pm 0.104$ & $Swift$/UVOT \\
59549.52 & 31.52 & $UVM2$ & $20.279 \pm 0.101$ & $Swift$/UVOT \\
59549.52 & 31.52 & $V$ & $18.060 \pm 0.152$ & $Swift$/UVOT \\
59550.32 & 32.32 & $r$ & $18.335 \pm 0.138$ & KAIT \\
59550.32 & 32.32 & $i$ & $18.454 \pm 0.112$ & KAIT \\
59551.31 & 33.31 & $r$ & $18.206 \pm 0.094$ & KAIT \\
59551.32 & 33.32 & $i$ & $18.291 \pm 0.102$ & KAIT \\
59553.93 & 35.93 & $UVW2$ & $20.753 \pm 0.132$ & $Swift$/UVOT \\
59553.93 & 35.93 & $U$ & $19.278 \pm 0.126$ & $Swift$/UVOT \\
59553.93 & 35.93 & $B$ & $18.509 \pm 0.114$ & $Swift$/UVOT \\
59553.93 & 35.93 & $UVM2$ & $20.699 \pm 0.123$ & $Swift$/UVOT \\
59553.93 & 35.93 & $V$ & $18.160 \pm 0.164$ & $Swift$/UVOT \\
59553.93 & 35.93 & $UVW1$ & $20.239 \pm 0.135$ & $Swift$/UVOT \\
59555.19 & 37.19 & $g$ & $18.651 \pm 0.032$ & SEDM \\
59555.20 & 37.20 & $i$ & $18.433 \pm 0.125$ & SEDM \\
59555.20 & 37.20 & $r$ & $18.354 \pm 0.045$ & SEDM \\
59564.21 & 46.21 & $i$ & $18.309 \pm 0.034$ & SEDM \\
59564.21 & 46.21 & $g$ & $18.985 \pm 0.051$ & SEDM \\
59564.21 & 46.21 & $r$ & $18.561 \pm 0.036$ & SEDM \\
59576.96 & 58.96 & $U$ & $19.806 \pm 0.106$ & $Swift$/UVOT \\
59581.40 & 63.40 & $U$ & $20.269 \pm 0.103$ & $Swift$/UVOT \\
59581.82 & 63.82 & $o$ & $18.809 \pm 0.075$ & ATLAS \\
59582.21 & 64.21 & $r$ & $19.033 \pm 0.081$ & SEDM \\
59582.21 & 64.21 & $g$ & $19.635 \pm 0.044$ & SEDM \\
59582.22 & 64.22 & $i$ & $18.749 \pm 0.032$ & SEDM \\
59582.22 & 64.22 & $g$ & $19.556 \pm 0.072$ & SEDM \\
59582.22 & 64.22 & $r$ & $19.099 \pm 0.607$ & SEDM \\
59582.23 & 64.23 & $i$ & $18.688 \pm 0.035$ & SEDM \\
59582.81 & 64.81 & $o$ & $18.758 \pm 0.168$ & ATLAS \\
59584.27 & 66.27 & $o$ & $18.947 \pm 0.081$ & ATLAS \\
59586.24 & 68.24 & $U$ & $19.989 \pm 0.101$ & $Swift$/UVOT \\
59586.26 & 68.26 & $o$ & $18.959 \pm 0.069$ & ATLAS \\
59590.25 & 72.25 & $o$ & $19.096 \pm 0.201$ & ATLAS \\
59591.75 & 73.75 & $U$ & $20.150 \pm 0.101$ & $Swift$/UVOT \\
59592.29 & 74.29 & $o$ & $19.202 \pm 0.165$ & ATLAS \\
59596.25 & 78.25 & $o$ & $19.299 \pm 0.180$ & ATLAS \\
59598.81 & 80.81 & $o$ & $19.002 \pm 0.147$ & ATLAS \\
59600.27 & 82.27 & $o$ & $19.237 \pm 0.133$ & ATLAS \\
59604.23 & 86.23 & $o$ & $19.231 \pm 0.110$ & ATLAS \\
59627.13 & 109.13 & $r$ & $19.800 \pm 0.147$ & ZTF \\
59628.12 & 110.12 & $r$ & $19.659 \pm 0.083$ & ZTF \\
59629.12 & 111.12 & $r$ & $19.648 \pm 0.085$ & ZTF \\
59630.12 & 112.12 & $r$ & $19.752 \pm 0.092$ & ZTF \\
59639.12 & 121.12 & $r$ & $19.506 \pm 0.181$ & ZTF \\
59640.12 & 122.12 & $r$ & $19.842 \pm 0.173$ & ZTF \\
59777.07 & 259.07 & $o$ & $19.446 \pm 0.309$ & ATLAS \\
59784.13 & 266.13 & $r$ & $21.385 \pm 0.231$ & LT \\
59784.13 & 266.13 & $i$ & $20.491 \pm 0.167$ & LT \\
59786.60 & 268.60 & $o$ & $20.332 \pm 0.228$ & ATLAS \\
59787.47 & 269.47 & $i$ & $20.338 \pm 0.190$ & ZTF \\
59788.44 & 270.44 & $r$ & $21.274 \pm 0.304$ & ZTF \\
59791.39 & 273.39 & $c$ & $21.061 \pm 0.312$ & ATLAS \\
59794.43 & 276.43 & $i$ & $20.380 \pm 0.248$ & ZTF \\
59795.10 & 277.10 & $g$ & $21.398 \pm 0.223$ & LT \\
59795.10 & 277.10 & $r$ & $21.496 \pm 0.271$ & LT \\
59795.10 & 277.10 & $i$ & $20.746 \pm 0.199$ & LT \\
59796.48 & 278.48 & $g$ & $21.489 \pm 0.287$ & ZTF \\
59809.42 & 291.42 & $g$ & $20.745 \pm 0.282$ & ZTF \\
59809.44 & 291.44 & $i$ & $20.480 \pm 0.240$ & ZTF \\
59809.47 & 291.47 & $r$ & $20.813 \pm 0.275$ & ZTF \\
59810.53 & 292.53 & $o$ & $20.244 \pm 0.275$ & ATLAS \\
59812.41 & 294.41 & $r$ & $21.236 \pm 0.259$ & ZTF \\
59813.04 & 295.04 & $o$ & $20.508 \pm 0.250$ & ATLAS \\
59815.40 & 297.40 & $i$ & $20.486 \pm 0.196$ & ZTF \\
59816.46 & 298.46 & $r$ & $21.254 \pm 0.288$ & ZTF \\
59816.46 & 298.46 & $g$ & $21.290 \pm 0.242$ & ZTF \\
59827.40 & 309.40 & $i$ & $20.665 \pm 0.255$ & ZTF \\
59828.45 & 310.45 & $r$ & $21.300 \pm 0.289$ & ZTF \\
59841.40 & 323.40 & $g$ & $21.802 \pm 0.298$ & ZTF \\
59846.34 & 328.34 & $i$ & $20.421 \pm 0.224$ & ZTF \\
59848.35 & 330.35 & $r$ & $21.279 \pm 0.274$ & ZTF \\
59850.34 & 332.34 & $r$ & $21.281 \pm 0.247$ & ZTF \\
59857.36 & 339.36 & $i$ & $20.458 \pm 0.273$ & ZTF \\
59872.47 & 354.47 & $c$ & $20.573 \pm 0.234$ & ATLAS \\
59883.28 & 365.28 & $i$ & $20.741 \pm 0.288$ & ZTF \\
59894.37 & 376.37 & $o$ & $20.312 \pm 0.298$ & ATLAS \\
59895.25 & 377.25 & $i$ & $20.546 \pm 0.300$ & ZTF \\
\enddata
\tablenotetext{a}{Rest-frame days since ATLAS $o$-band peak on MJD 59518.0.}
\end{deluxetable}

%% file: appendix/SN2021aaev_spectroscopy_table.txt
\begin{deluxetable*}{lcccccc}[h]
\tablecaption{Spectroscopic Observations of SN~2021aaev.\label{tab:spectroscopy}}
\tablehead{
\colhead{UT} & 
\colhead{MJD} & 
\colhead{Phase\tablenotemark{a}} & 
\colhead{Telescope} & 
\colhead{Exposure} & 
\colhead{Grism(s),} & 
\colhead{Wavelength Range} \\
\colhead{} & 
\colhead{} & 
\colhead{[day]} & 
\colhead{+Instrument} & 
\colhead{[s]} & 
\colhead{slit} & 
\colhead{[\AA]}
}
\startdata
2021-10-10 & 59497.96 & -20.04 & LT+SPRAT & 600 & - & 4000--8000 \\
2021-10-11 & 59498.15 & -19.85 & SEDM+P60 & 2700 & - & 3800--9200 \\
2021-10-12 & 59499.17 & -18.83 & NOT+ALFOSC & 1800 & Gr\#4, 1.3" & 3500--9600 \\
2021-10-14 & 59501.29 & -16.71 & VLT+X-Shooter & 1200/1229/300 & Gr\#4, 1.3" & 3000--24800 \\
2021-10-16 & 59503.33 & -14.55 & NTT+EFOSC2 & 600 & Gr\#13, 1.0" & 3650--9250 \\
2021-10-16 & 59503.34 & -14.65 & SEDM+P60 & 2700 & - & 3800--9200 \\
2021-10-22 & 59509.22 & -8.78 & SEDM+P60 & 2700 & - & 3800--9200 \\
2021-10-26 & 59513.10 & -4.90 & NTT+EFOSC2 & 1800*2+1800 & Gr\#11+Gr\#16, 1.0" & 3350--10000 \\
2021-11-06 & 59498.15 & 6.15 & SEDM+P60 & 2160 & - & 3800--9200 \\
2021-11-27 & 59545.19 & 27.19 & NTT+EFOSC2 & 1800*2 & Gr\#11, 1.0" & 3350--7470 \\
2021-12-03 & 59551.17 & 33.17 & NTT+EFOSC2 & 1800*2 & Gr\#16, 1.0" & 6000--10000 \\
2021-12-22 & 59570.91 & 52.91 & NOT+ALFOSC & 3600 & Gr\#4, 1.0" & 3500--9600 \\
2022-01-03 & 59582.09 & 64.09 & NTT+EFOSC2 & 2700+2700 & Gr\#11+Gr\#16, 1.0" & 3350--10000 \\
\enddata
\tablenotetext{a}{Rest-frame days since ATLAS $o$-band peak on MJD 59518.0.}
\end{deluxetable*}

%% file: appendix/MCMC.txt
\begin{figure*}[h]
    \centering
    \includegraphics[width=\linewidth]{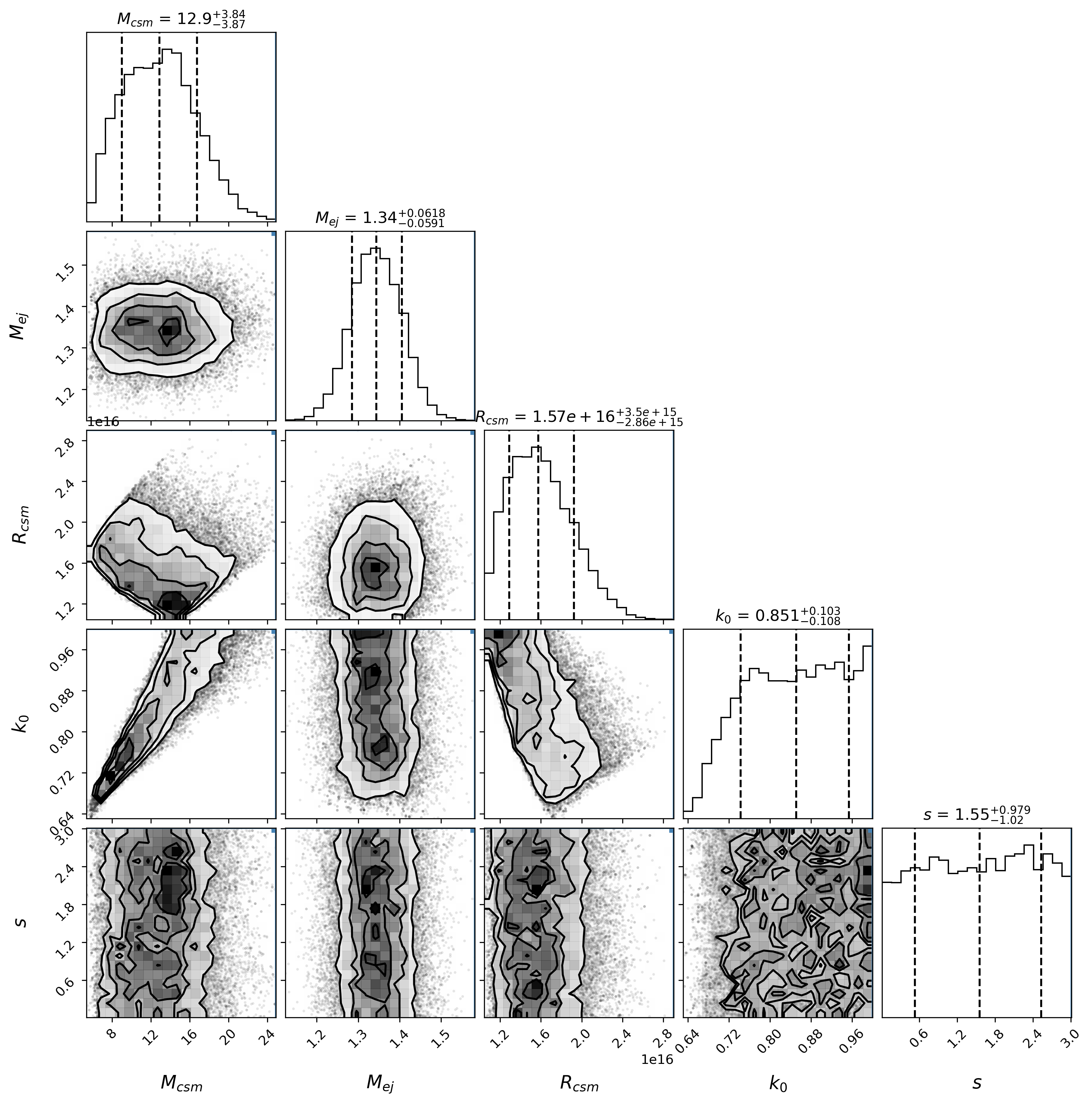}
    \caption{Corner plot of fitted parameters of SN\,2021aaev using the ``interior breakout'' CSM-interaction analytical model \citep{khatami2024}, showing their median values and $1\sigma$ spread. The sample was drawn using \texttt{emcee}.}
    \label{fig:Khatami_Kasen_parameters}
\end{figure*}

\begin{figure*}[h]
    \centering
    \includegraphics[width=\linewidth]{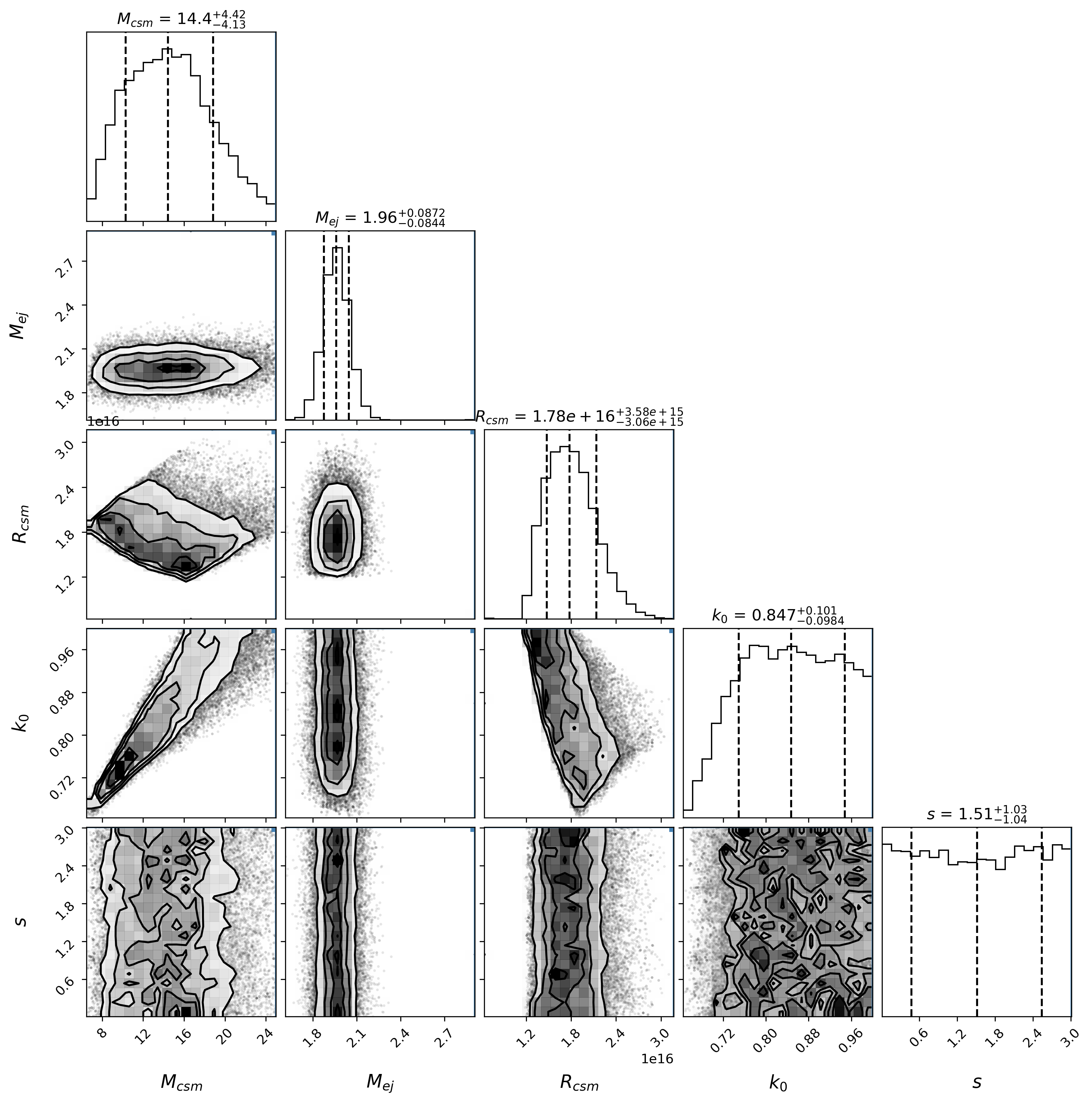}
    \caption{Corner plot of fitted parameters of SN\,2021aaev using the ``interior breakout'' CSM-interaction analytical model \citep{khatami2024}, showing their median values and $1\sigma$ spread. This has the additional constraint that the energy conversion efficiency is $\epsilon=0.6$. The sample was drawn using \texttt{emcee}.}
    \label{fig:Khatami_Kasen_parameters2}
\end{figure*}